# Highly Efficient Hydrogen Storage of Sc Decorated Biphenylene Monolayer near Ambient-temperature: An Ab-initio Simulations

*Mukesh Singh[a], Alok Shukla[a*], Brahmananda Chakraborty[b,c]*

[a]Department of Physics, Indian Institute of Technology Bombay, Powai, Mumbai 400076, India

[b]High Pressure and Synchrotron Radiation Physics Division, Bhabha Atomic Research Centre, Trombay, Mumbai, India

[c]Homi Bhabha National Institute, Mumbai, India

**Abstract:** The energy demands for the growing development of society need to be catered with alternative and green fuels like hydrogen energy for a lasting and sustainable culture. One essential component of the hydrogen economy is the efficiency of its storage. We have studied the hydrogen-storage capability on a recently synthesized Biphenylene (BPh) decorated with Sc using the first-principles density functional theory (DFT) and ab-initio molecular dynamics (AIMD) techniques. Scandium attaches BPh sheet strongly with binding energy -3.84 eV, and single Sc decorated on BPh can absorb a maximum of five $H_2$ molecules resulting in a high gravimetric weight percentage of 11.07, which is significantly higher than DoE's ultimate criteria (6.5 wt%). Using van't Hoff equation, strongly and weakly attached hydrogens correspond to desorption temperatures of 200 K and 397 K with an average of 305 K. The high binding of Sc to BPh is due to charge donation of 3d orbital of Sc to 2p orbital of C. The interactions between absorbed $H_2$ and BPh+Sc are due to charge transfer from 3d-orbital of Sc to σ* bond of $H_2$ molecules and backdonation from σ bond of $H_2$ to empty 3d-orbital of Sc known as Kubas type interaction. Furthermore, phonon and AIMD simulation confirm BPh+Sc stability, and the presence of an energy barrier shows no probability of Sc-Sc clustering on BPh. So theoretically stable BPh+Sc showing high gravimetric weight percentage with an average 305 K desorption temperature, might be a potential candidate for solid-stage hydrogen devices.

---

*Corresponding author: Prof. Alok Shukla
E-mail address: shukla@phy.iitb.ac.in
Phone Numbers: +91-22-2576-7576

## 1. Introduction

Rapidly evolving modern technologies enhance the dependability of energy resources, particularly on conventional fuels. Energy Information Administration (EIA) estimated 28% growth in energy consumption between 2015 and 2040 [1]. Energy records for 2020 show that approximately 80% of energy is extracted from fossil fuels [2]. Considering fossil fuel demands and their exploitation, Shafiee et al. have predicted using Klass modified formula that gas, oil, and coal would be exhausted in the next hundred years [3]. Hence, there will be an immediate requirement for renewable, green, abundant energy resources to protect the environment from further deterioration. To fulfill energy demands, DOE Department of Energy (DoE) has specified criteria for suitable, efficient, scalable, renewable energy sources such as hydrogen fuel, biomass, and their storage [4–10]. DoE mentions hydrogen energy as one of the most beneficial alternatives because of the following reasons. First, $H_2$ fuels (120 MJ/kg) contain approximately three times more energy density than gasoline (44 MJ/kg) [11]. Second, it is abundant in the universe and in the form of $H_2O$ on earth. Third, on utilization, it produces pollution-free energy and $H_2O$ as a byproduct. Interestingly, $H_2O$ can be electrolyzed to produce $H_2$ making the whole process cyclic and $H_2$ never lasting energy resource. For a suitable, scalable solid-state hydrogen material, US DoE has specified two main criteria: (i)- the binding energy of the attached hydrogen molecule must range between 0.2-0.7 eV/$H_2$, (ii) gravimetric weight percentage of $H_2$ on the system must be more than 5.5 wt% and 6.5 wt% according to 2025 target and ultimate target [12]. Several technics have been used to store hydrogen energy in various materials. The primary way to store hydrogen energy is to compress and stock $H_2$ in a tank [13], known as compressed gas hydrogen storage. However, it requires a large vessel with a specially created carbon-reinforced fiber tank to consider the concern of safety issues[14]. These unique walls and large vessel requirements make it expensive and unsuitable for transport vehicles. Another method of cryogenic liquified hydrogen gas storage has been considered to avoid high compression. But producing cryogenic containers and maintaining temperatures lower than the boiling temperature (20 K) of $H_2$ is costly as well [15,16]. So in practice, cryo-compressed $H_2$ ($CCH_2$) is utilized because it requires moderate conditions of temperature and pressure that can be created at a relatively low cost. However, $CCH_2$ also requires special attention for safety and is costly for everyday practical usefulness. So alternative materials are constantly being explored and characterized by their ability to adsorb/absorb hydrogen. They are mainly found as hydrides [17], porous materials [18], and carbon nanomaterials [19] which we will discuss briefly. Hydrogen absorbing materials and its container to store $H_2$ for application purposes are called solid-state hydrogen storage devices. In these devices, hydrogen gas is absorbed in compact, safe forms, which might make them future hydrogen storage devices for onboard and industrial applications [20].

**1.1 Hydride hydrogen storage:** Hydrogen storage materials in solid-state hydride devices are mainly elemental hydride, metal alloy, and complex hydride. Elemental metals hydrides of Li, Mg Ca, Al, etc., are useful due to their high hydrogen density and lightweight. Additionally, the low resistivity and recyclability of MgH$_2$ make it the main focus of elemental hydride [21,22]. However, sluggish kinetics and thermodynamic reactions of adsorption and desorption are major disadvantages of elemental hydride [23]. Metal alloy/complex hydrides of $A_xB_yH_z$ form such as Mg$_2$NiH$_4$, FeTiH$_{1.95}$, LaNi$_5$H$_{6.7}$, ZrMn$_2$H$_{3.6}$, are reported to have gravimetric weight percentages of 7.6, 3.6, 1.89, 1.37, 1.75, 2.10 wt%, respectively [24]. However, metal alloy/complex hydrides are not suitable for onboard hydrogen storage because of their heavy weight, slack kinetics, low reversibility, and high dehydrogenation temperature [25].

**1.2 Porous material hydrogen storage:** High porosity materials such as metal-organic frameworks (MOFs), Zeolites, and their related composite materials are described with the chi-hypothesis, which suggests large surface area and high porosity mainly contribute to physisorption for H$_2$ molecules [26] making it fast and highly reversible [27]. Materials having large surfaces and porous materials are predicted to contribute to high hydrogen storage with lower desorption temperatures. Here we will briefly highlight important works of these materials with their advantage and drawback. MOFs defines as materials with organic ligands attached to inorganics building units (metal ions, clusters). Over two decades ago, MOFs were investigated as potential hydrogen storage materials [28]. Recently, Chen et al. have reported metal trinuclear-based MOFs such NU-1500-Al, which show impressively high gravimetric and volumetric capacity (14.0 wt %, 46.2 g/l) under swing conditions of temperature and pressure between 77 K/100 bar → 160 K/5 bar [29]. However, MOFs as onboard hydrogen storage materials are still challenging because of their low gravimetric, volumetric capacity, cost, etc. [18]. Zeolites are aluminosilicate materials constructed from relatively heavier elements [18]. Lagmi et al. synthesized different possible forms of geometry and pores, namely A, X, Y, and RHO type zeolite, and reported H$_2$ storage capacity of 1.79 wt%, 1.81 wt%, and 1.74 wt % at 77K and 15 bar for NaX, NaY and MgY respectively [30] which are relatively low with other storage materials. However, with computational simulations, high capacity zeolite templated carbon material have reported [31,32] whose experimental verifications are needed.

**1.3 Carbon nanomaterials:** Similar to porous materials, pristine carbon-based materials physisorb hydrogen, resulting in a low gravimetry efficiency [33]. Ma. et al. have verified the hydrogen adsorption behavior on graphene sheets and reported hydrogen intake of only 0.4 wt% and < 0.2 wt% at cryogenic and room temperatures, respectively [34]. Similar low efficiency of 0.1 wt% (298K under 10 bar) and 1.4 wt% (298 K under 50 bar) have been observed for graphene nanosheet and graphene oxide [35,36], respectively. Due to low wt%, pristine materials in which H$_2$ molecules interact with physisorption are not suitable for H$_2$ storage at ambient temperature. To address the low gravimetric wt%, researchers raise the interaction between H$_2$ and absorbate material. Many studies have reported effective ways to improve the hydrogen storage capability of carbon-nanomaterial by decorating them with alkali, alkaline, or transition metal. Metal functionalized Fullerenes have been outlined to enhance gravimetric weight percentage up to ~6-9.5 wt% [37–43]. Similar trends of increasing wt% are reported for metal decorated carbon nanotube (SWCNT) [44–47], metal decorated on pristine graphene [48–52], and metal decorated on B/N doped graphene [53–57]. Metal decorated 2D nanomaterials with triple bonded such as Y+graphyne, Li+Holey graphyne, etc., have also been reported to boost gravimetric weight percentage [58,59]. The increased wt% of metal decorated carbon nanomaterials are contributed by Kubas type [60] and spillover mechanism [61], whose strength lies between physisorption and chemisorption. Modak et al. have studied the contribution of decorated transition metals (Y, Zr, Nb, Mo) on carbon nanomaterials (SWCNT) and discussed that transition metals with a less number of d-electron are more efficient for decoration [62]. Experimental works on hydrogen storage have also confirmed the enhancements in gravimetric wt% of metal decorated carbon nanomaterials. For instance, Mg nanocrystal condensed in graphene oxide was observed to have 6.5 wt% and 0.105 $H_2$ per liter [63]. Gu et al. in 2019 reported gravimetric hydrogen uptake of 5.7 wt% at 473K [60] for Ni and Al codoped graphene [64]. For Ni, Mg composite of graphene, Tarasov et al. have reported $H_2$ gravimetric wt% of more than 6.5 [65]. Exploiting spillover mechanism, Samantaray et al. have reported ~4.5 wt % at 30 bar and 25°C in $Pd_3Co$ functionalized boron-doped graphene [66].

Recently, a new carbon allotrope biphenylene (BPh) has been synthesized [63] with properties of high stability and mechanically robust [67]. In this work, we have explored the utilization of this new material for onboard lightweight vehicles as hydrogen storage. Denis et al. have studied alkali metal (Li) decorated hydrogen on BPh [68], but Li binds weakly to BPh sheet. Recently, Ca, K, and Zr decorated Biphenylene hydrogen storage works have been reported in which Ca, K, similar to Li, binds weakly, and Zr atom is heavier than Sc [69,70]. So we consider firmly attached transition lightest metal (Sc) decorated BPh for our simulation for high gravimetric wt %. Transition metal Sc is considered because fewer d-electron transition metals are more efficient for $H_2$ adsorption [62]. First, we checked the stability of BPh+Sc with energy, phonon, and AIMD, as described by Malyi et al. [71]. Then we performed DFT computation with PBE functional, including van der Waals interactions for the Sc-decorated BPh (BPh+Sc). Our DFT simulations predicted that BPh+Sc sheet adsorbs up to five $H_2$ molecules resulting in 11.07 wt%. Using van't Hoff equation, the average desorption temperature is calculated with 305 K. We have discussed adsorption mechanisms by the use of PDOS, Bader charge analysis, charge density plots, and Kubas-type interactions. BPh+Sc has a sufficiently high energy barrier between possible diffusion paths, leading to negligible diffusion and clustering. Predicting the stability of the system, the presence of energy barriers, and high wt% of adsorbed H$_2$, we strongly believe that BPh+Sc system can be a suitable material for high-capacity reversible solid-state hydrogen storage.

## 2. Computational Methods

The DFT calculations are performed using VASP DFT code which has implemented PAW method [72,73]. Generalized gradient approximations are used for all computation (GGA-PBE) [74] in which exchange-correlation under binds the adsorbed atom/molecules contrary to local density approximation (LDA), where exchange-correlation overbinds adsorbed atom/molecules [75–77]. To include weak van der Waals interaction, we have corrected GGA calculated results with Grimme-D2 methods [78]. The pseudopotential of C, Sc, and H have considered electronic configurations $2s^22p^2$, $3p^64s^23d^1$, and $1s^1$ as valence electrons to perform geometry relaxation, single point self-consistent field, and non-self consistent field calculations. The energy convergence criteria for an electronic loop and force convergence for an ionic loop were chosen to be $10^{-6}$ eV and $10^{-2}$ eV/Å, respectively. Energy cutoff of 500eV for plan waves and k-points of 7x7x1 for Brillouin zone sampling in the Monkhorst scheme [70,79] were selected. For geometry relaxation, the conjugate gradient algorithm is used. For thermal stability, phonon calculations are performed using vaspkit (generated K-path for phonon spectrum) [80], phonopy [81], and VASP with consideration of 2x2x1 supercell, 9x9x1 K-points grid, 800 eV energy cutoff, $10^{-3}$ eV/Å force convergence, $10^{-9}$ eV energy convergence criteria. After gravimetric wt%, practical feasibilities are checked by the calculation to avoid metal-metal and high temperature stability. The diffusion energy barrier is calculated self consistently on multiple points along the path, joining the most stable (initial) position and possible diffusion position (final) of Sc [82]. For high-temperature stability of the system, ab-initio molecular dynamics (AIMD) calculations are performed [83] in two steps using VASP, which implements fourth-order Predictor–corrector algorithms to integrate equation of motion [84]. First, in a microcanonical ensemble, the temperature is raised from absolute zero to the desired temperature in time steps of 1fs. Next, the system is observed for 5 ps in the canonical ensemble at a constant temperature provided by the Nose-Hoover thermostat [85].

## 3. Results and Discussions

**3.1 Biphenylene structure:** First, a unit cell of Biphenylene (BPh) is modeled such that its chemical structure overlaps with STM images of experimental works [86]. We optimized a 2x2x1 supercell of BPh structure with lattice constant a=4.52Å and b=3.76Å of the rectangular unit, which has space group Pmmm [67]. The unit cell of BPh are marked with magenta color in plots of supercell in Fig. 1. Similar to graphene, carbon atoms in BPh are sp$^2$ hybridized but with different bond lengths and angles, as shown in Fig. 1. The bond lengths of the optimized structure are 1.45Å,

1.40Å, 1.45Å, and bond angles are $\theta_1$=90.00º, $\theta_2$=109.89º, $\theta_3$=125.05º as indicated in Fig. 1 perfectly matching with the previous literature (1.45Å, 1.40Å, 1.45Å, 90º, 109º, 125º) [67,87]. The bond lengths in BPh are close in value to those in graphene, suggesting high structural stability of its structure [88]. For thermal stability, phonon frequency calculations are performed, and all frequencies positive ensure the thermal stability of our modeling of BPh Fig. 2.

**3.2 Decoration of Sc on BPh**

On DFT optimized BPh structure, we explored all possible positions to decorate over it with Sc transition metal. We have chosen Sc because it has a lesser number of d-electrons (1e), leading to high hydrogen adsorption. We have considered nine different positions to decorate Sc: two on top of carbon atoms (A1, A2), four on top of the middle points of bonds (B1, B2, B3, B4), and three on top of the center of pores (C1, C2, C3) as plotted in Fig. 3. Let us denote Sc doped at position P on BPh with BPh+Sc(@P), i.e., BPh+Sc(@A1) denotes Sc doped at A1 on BPh; BPh+Sc(@A2) denotes Sc doped at A2 on BPh and so on. The adsorption energy of Sc doped on BPh, $E_b(Sc)$, is calculated by the following formula:

$$E_b(Sc) = E_{BPH+Sc} - E_{BPH} - E_{Sc} \qquad (1)$$

Where $E_{BPh+Sc}$, $E_{BPh}$, $E_{Sc}$ are ground state energies of BPh+Sc, BPh, and isolated Sc atom, respectively.

In Fig. 3, we have plotted Sc-decorated structure before and after optimization where tail and head of an arrow point to initial and final positions of decorated Sc atom. From Fig. 3, we found that after relaxation, Sc decorated on C1, C2, C3, and B1 remained at their initial positions, while Sc decorated on B2, B4, and A1 moved to B2', Sc doped on B3 moved to near C1, A2 moved to B1. Finally, we tabulated the binding energy of Sc-decorated BPh sheet and named them according to their final position. From Table 1, we can conclude three things: First, binding energies of Sc doped at different positions on BPh sheet are different because of different environments around the decorated Sc atom. Second, bindings of Sc on BPh sheet at octagon, hexagon, and square positions are much greater than that of the binding energy previously reported for Li decorated on BPh (-1.59, -1.53, -1.40 eV) [68]. Hence, Sc is more firmly attached to BPh, so it prevents detachment at higher temperatures rather than Li. These results are similar to previously reported transition metal decoration and are stronger than that of alkali/alkaline metals. Third, B2' and C2 are found to be the first and second most stable decorated positions for Sc on BPh sheet. Therefore, all the analyses of hydrogen storage capacity are performed for B2' and C2 decorated Sc on BPh.

**3.3 Binding mechanism of Sc with BPh sheet**

First, we computed the phonon spectrum of Sc decorated BPh sheet to check the thermal stability and plotted its dispersion in Fig 2. Fig. 2 shows all the frequencies are positive, indicating thermal stability of BPh+Sc sheet. Interestingly, we noted that phonon-spectrum of BPh+Sc (green color) on decorating Sc heavier atom (than C-atom), frequency near fermi energy, and highest optical mode shifted lower as reported by previous literature [89,90]. After stability confirmation, we analyze the interaction of Sc atom decorated at B2' and C2 by the use of density of states analysis, Bader charge analysis, and charge density plots.

**Density of states analysis:** The electronic structures such as density of states are good descriptors to represent the interaction between surface and adsorbate [91]. The total density of states of BPh before and after Sc atom decoration are plotted in Fig. 4, which shows that after decoration, the change in the profile of density of states indicates a strong interaction between Sc and BPh sheet. In Fig. 5, the partial density of state for 2p orbital of carbon atom in BPh+Sc and BPh are plotted. Fig. 5 shows that after decorating Sc on BPh, states near -1 eV have increased, indicating that after decorating Sc, the carbon atoms of BPh have gained some charges. This charge gain must be donated from some atoms. In Figs. 5 (c, d), we have plotted 3d orbital density of states of scandium atom for BPh+Sc and isolated Sc. On comparing Figs. 5(c) and (d), we see that the population of states at the Fermi level has decreased greatly after decorating Sc on BPh, indicating that Sc atom has donated the charge. In conclusion, the Sc atom has lost some charge, and BPh sheet has gained, which can also be thought of as some charge has been transferred from Sc atom to carbon atom.

**Bader charge analysis:** Partial charge density analysis indicated the charge transfer qualitatively. We analyzed Bader charges [92] of Sc decorated BPh sheet at every possible position for quantitative analysis and presented the results in table 1. At B2' position, the charge transfer from Sc atom to BPh sheet and distance between Sc to the nearest carbon of BPh sheet are 1.34e and 2.15Å. This amount of charge transfer and short distance leads to strong binding of Sc to BPh. For C2 position, in comparison with B2' position, charge transfer are smaller (1.29 e) and distance (2.22 Å) are larger, indicating smaller binding energy at C2 than B2' position.

**Charge density plots:** To picture the charge transfer of systems BPh+Sc(B2') and BPh+Sc(@C2), we plotted the difference in charge density, $\rho(BPh+Sc@B2')-\rho(BPh)$ and $\rho(BPh+Sc@C2)-\rho(BPh)$ in Figs. 6 and S1, respectively. The red color denotes the lost charge region, and Blue color denotes the gained charge region. In Fig. 6, the dark blue region nearest two carbon atoms of Sc has gained more charges from decorated Sc atom. Similarly, in Fig. S1 the nearest two carbon atoms have gained more charges.

**3.4 Adsorption of Hydrogen molecules on Sc decorated BPh sheet:**

After finding a strongly binding Sc atom on BPh sheet, we proceed towards adsorption of $H_2$ molecules on BPh+Sc system. We put one hydrogen molecule at 2.24 Å away on the top of Sc atom of BPh+Sc system. After geometry optimization BPh+Sc+1$H_2$, $H_2$ molecule is at a distance of 2.42 Å from Sc, and H-H bond of absorbed $H_2$ molecule increases from 0.74 Å to 0.76 Å. The binding energy of $H_2$ doped on Sc decorated BPh sheet can be calculated by the following formula [66,74]:

$$BE_{H_2} = E_{BPH+Sc+H_2} - E_{BPH+Sc} - E_{H_2} \qquad (2)$$

Where $E_{BPh+Sc}$, $E_{BPh}$, $E_{H2}$ are DFT ground state energy of BPh+Sc+1$H_2$, BPh+Sc, and isolated $H_2$ molecules, respectively.

The binding energy of the first hydrogen molecule on BPh+Sc system turns out to be -0.34 eV/$H_2$. We repeated the calculation with Grimme-D2 correction to include the van der Waals correction, and the corrected binding energy is found -0.42 eV/$H_2$. The bond length of adsorbed hydrogen molecules increases without dissociating $H_2$ because it gains some charge through Kubas-type interactions [93,94]. Now we doped $2^{nd}$ and $3^{rd}$ hydrogen molecules at a distance of 2.4 Å over Sc such that it is at maximum distance from the previous doped $H_2$ molecules. After optimization, the distance of $2^{nd}$ and $3^{rd}$ hydrogen molecules was 2.22 Å with an average van der Waal corrected binding energy of -0.53 eV/$H_2$. The average binding energy of the last two adsorbed hydrogen molecules is calculated using the following general formula:

$$BE_{H_2} = \frac{1}{n}\left[E_{BPH+Sc+(m+n)H_2} - E_{BPH+Sc+mH_2} - nE_{H_2}\right] \qquad (3)$$

Where $E_{BPh+Sc+(m+n)H2}$, $E_{BPh+Sc+mH2}$, $E_{H2}$ are ground state energy of BPh+Sc+(m+n)$H_2$, BPh+Sc+m$H_2$, and isolated $H_2$ molecules, respectively.

The bond length of H-H atoms in the last two doped hydrogen molecules increases from 0.74Å to 0.779 Å. Again, we doped two hydrogen molecules over Sc atom at 2Å distance away, considering $4^{th}$ and $5^{th}$ hydrogen molecules maintain maximum distance from the previous doped hydrogen molecules. After relaxation, furthermost $H_2$ molecules are found to be at a distance of 3.682 Å with van der Waal corrected average adsorbed binding energy -0.272 eV, and bond length increasing from 0.74 Å to 0.752 Å. We repeated this process as long as the adsorption energies of $H_2$ molecules were within the range of DoE criteria. We have found that BPh system can adsorb up to five $H_2$ molecules on each Sc atom doped on B2' positions and tabulated their energy and distance from the metal in table 2. The changing patterns of adsorption energy of $H_2$ molecules on BPh+Sc system are discussed in the interaction mechanism section (next section). All the geometrically relaxed structures are plotted in Fig. 7. Also, we repeated the hydrogen adsorption process for Sc decorated at C2 position, plotted the optimized structure in Fig. S2, and tabulated their binding energies in table S1. Considering our results and DoE's criteria, we conclude that five $H_2$ molecules should be considered for hydrogen storage on Sc decorated BPh sheet.

**3.5 Interaction of $H_2$ molecules with BPh+Sc:** To understand the interaction between H2 molecules and Sc decorated BPh sheet, we have again investigated the partial density of states, Bader charges analysis, and charge density difference plots. In Fig. 8 (a,b), we have plotted PDOS of 3d orbital of Sc before and after adsorption of hydrogen molecule on BPh+Sc system. On comparing Figs. 8 (a) and (b), we can observe that near -1.5 eV, some states are depleted. Hence, during the process of $H_2$ adsorption, some charges have been lost by Sc atom. To find where the lost charge of Sc atom gets transferred, we have compared PDOS of 1s of adsorbed $H_2$ molecules in BPh+Sc+1$H_2$ and 1s orbital of H of isolated $H_2$ molecule in Fig. 8 (c,d). On comparing Fig. 8(c) and (d), we notice that enhancement of states near -1.5eV in 1s orbital of absorbed $H_2$ molecules, indicating some charges have been gained by adsorbed $H_2$ molecules. The charge distribution surrounding adsorbed $H_2$ molecules can be seen from charge density difference plots. In Fig. 9, we have plotted charge density difference of BPh+Sc+1$H_2$ to BPh+Sc and isolated $H_2$ molecules, $\rho(BPh+Sc+1H_2)-\rho(BPh+Sc)-\rho(H_2)$ with isovalues of $3.5\times10^{-3}$e where cyan and yellow color represent the charge gain and charge lost regions. From Fig. 9, we can see that some charges has been lost from bonding-state (σ-bond), and some charge has been gained to anti-bonding (σ*-states) as previously explained in literature [95]. Quantitatively, Bader charge analysis computes an overall charge gain (0.046e) due to charge transfers from σ and σ*-bonds of absorbed $H_2$. From these analyses of PDOS, Bader analysis, and charge difference plots, we found after adsorbing $H_2$ molecules on BPh+Sc, $H_2$ molecules gain some charges due to electron transfer from the 3d orbital of Sc to σ* bonds (cyan color in Fig. 9) $H_2$ molecules and some charge are donated from σ-bonds (yellow color in Fig. 9) of $H_2$ molecules. This kind of charge transfers from 3d orbital to σ*-bonds of $H_2$ molecules, and back-donation of some charge of $H_2$ molecules (from σ-bonds) is called Kubas-interaction [93,94]. After adding 3$H_2$, and 5$H_2$ molecules, the charge maximum gained by adsorbed hydrogen atoms are 0.068e, 0.042e, and their enlarged H-H bond lengths are noted to be 0.779Å and 0.752Å [96]. Hence, considering effectiveness of Kubas interaction up to 5$H_2$ molecules, we have calculated the gravimetric weight of hydrogen storage by considering 5$H_2$ adsorbed on each Sc, which is 11.07 wt%. Finally, The hydrogen storage gravimetric weight percentage of previous works of BPh like structures, such as carbon allotropes, B/N doped 2D nanomaterials, are compared in table 3.

**3.6 Gravimetric weight percentage (wt%) and desorption temperature:**

B2' decorated Sc atom on BPh are most stable and adsorb the maximum number of hydrogen molecules, i.e., five $H_2$ molecules. We consider Sc atom decoration on B2' position of each hexagon in BPh sheet to get high gravimetric wt%. Our simulation considered a supercell containing 2x2x1 unit cells (6C atoms/unit cell), indicating we can decorate four Sc atoms on each side, i.e., 24C+4Sc Fig. 10(a). Considering BPh is a 2D sheet, both surfaces can be decorated with Sc atom resulting in 8 Sc per simulation cell ie. 24C+8Sc Fig. 10(b). Above DFT simulations show that each Sc atom adsorbs up to $5H_2$ molecules. Hence, one side decoration of 4 Sc atoms results in $20H_2$ absorbed on BPh (24C+4Sc+$20H_2$). Similarly, both side decorations of 8 Sc atoms result in $40H_2$ absorbed on BPh (24C+8Sc+$40H_2$) of which geometrically optimized structure are plotted in Fig. 7(d). Gravimetric weight percentages of 7.93% and 11.07% are evaluated for one-sided and both-sided Sc-decorated BPh sheets, respectively. $H_2$ Gravimetric weight percentage is calculated using [97]:

$$H_2 wt\% = \frac{n_{H2} * w_{H2}}{n_C * w_C + n_{Sc} * w_{Sc} + n_{H2} * w_{H2}} \qquad (4)$$

Where $n_C$, $n_{Sc}$, $n_{H2}$ are the numbers of C, Sc, and $H_2$ atom/molecules in unit cell, and $w_C$, $w_{Sc}$, $w_{H2}$ are weights of C, Sc and $H_2$.

In order to desorb hydrogen molecules from solid-state storage BPh+Sc for utilization, desorption temperature calculation is required, defined as the temperature at which hydrogen molecules detached from BPh+Sc system. From table 2, $3H_2$ and $5H_2$ adsorbed molecules with binding energy of -0.53eV and -0.27eV corresponds to 6.95 and 11.07 wt% computed with equation (4). The average binding energies of absorbed $H_2$ molecules with van der Waals correction is -0.412 eV and its corresponding desorption temperatures of this system are 305 K, calculated with the following van't Hoff equation [98–100].

$$T_d = \frac{E_b}{k_B}\left(\frac{\Delta S}{R} - \ln P\right)^{-1} \qquad (5)$$

Where $T_d$, $E_b$, $k_B$, P, R, and $\Delta S$ are desorption temperature, adsorption energy, Boltzmann constant, ambient pressure, gas constant, and change in entropy for $H_2$ molecules going from gas phase to liquid phase (130.68 $JK^{-1}mol^{-1}$ [101]), respectively. The strongly (-0.53eV) and weakly (-0.27eV) attached hydrogen molecules set the maximum and minimum limit of desorption temperature. Using von't Hoff equation, we found the desorption temperature of our system lie between 200-397K with the average desorption of 305K. Finally, to see desorption at room temperature, we have performed AIMD of five $H_2$ molecules attached to each Sc decorated at every hexagon of BPh in two steps. We first raised the temperature of system in microcanonical ensemble from 0 to 300K with steps of 1fs for 5ps. Then in second states, using Nose-Hoover thermostats, we kept the temperature constants at 300K for 5ps and plotted the final results in Fig. S3. We have considered $H_2$ molecules desorbed to our system if maximum distance of metal-$H_2$ is at larger distance than 4.26 Å as in literature [102] and black dotted box marked 4.26 Å distance from over and below of decorated Sc atoms. AIMD simulation of 2x3x1 super cell of BPh, i.e., 36C+12Sc+60H$_2$, shows at room temperature ~23% H$_2$ molecules get desorption from the system and 77 % remains attached to system.

**3.7 Diffusion barrier:** To realize experimentally above calculated hydrogen storage wt%, hydrogen storage materials (BPh+Sc) must ensure no metal-metal clustering confirmed by energy barrier calculation. We first check clustering by considering simply free Sc particles if they cluster, then we consider decorated Sc atoms on BPh. Since cohesive energy (3.90) [103] of bulk Sc is more than binding energy (3.83), free Sc atoms will cluster and forms Sc bulk [82]. Hence we consider calculating energy barrier between the most stable point (B2') and its neighboring points (C3)[100]. We performed single point calculation at five different points between B2' and C3 positions and plotted the relative energy with respect to the distance Fig. 12. The calculated energy barrier of 3.48 eV is much higher than thermal energy kinetic energy (0.02 eV) at room temperature using 1.5k$_B$T indicate negligible diffusion and hence no clustering.

**3.8 AIMD simulations:** All the above DFT calculations are performed at 0 K, but hydrogen storage devices are intended to be used at ambient temperature. We performed AIMD to check room temperature stability in two steps: **1-** we raised the temperature of BPh+Sc to room temperature for 5ps with the steps 1fs, keeping every step in a microcanonical ensemble (NVE). **2-** Using the final structure of the first step and kept the system at constant room temperature with Nose-Hoover thermostats for 5ps and observed the evolution of structures. We have repeated AIMD simulations for different configurations of Sc atoms decorated on BPh at different temperatures. After 2$^{nd}$ step of AIMD, the final structure and bond length variations are plotted in Figs. 13-16 and S4. Fig. 13(a) shows that structure composite (BPh+Sc) remains stable at 300K. Fig. 13(b) shows a negligible distance variation between Sc to its nearest carbon. We also performed AIMD for multiple Sc atoms decorated at B2' positions of BPh (BPh+4Sc) at 300K and at high temperature (~ 700K) for stability Fig. 14. Finally, stability of Sc atoms decorated on both sides of BPh sheet at each B2' positions are checked and found with AIMD Fig. 15. The variations of separation between two Sc and other Sc-Sc distances are plotted in Fig. 16 and S4, respectively. From these negligible variations of distances between two atoms (C-C, C-Sc, and Sc-Sc), we conclude that Sc decorated at each hexagon of BPh structures on one side and both sides are stable.

We have also explored the possibility of simultaneously decorating Sc atoms at different sites with the help of optimization and AIMD simulations. First, we considered Sc decoration on all porous sites, i.e., square, hexagon, and octagon decoration of Sc, but both geometrical relaxation and AIMD-simulations resulted in distorted structures Fig. S5. From these, we conclude that Sc cannot be decorated in all possible porous sites. Again, we consider decorating Sc atoms at only octagon and hexagon, and the structure gets distorted Fig. S6. Hence our simulations (DFT+AIMD) show that Sc decoration should be considered only at hexagonal positions for hydrogen storage.

**3.9 Summary and Conclusions**

DFT simulations show **1.** The Sc binds strongly at B2' and C2 positions on biphenylene with binding energies -3.83 eV and -3.56 eV, respectively. **2.** BPh+Sc@B2' can adsorb up to five $H_2$ molecules resulting in a high gravimetric weight percentage of 11.07%. **3.** The average binding energies of absorbed hydrogen molecules are -0.373 eV and -0.412 eV for DFT and DFT+D2. **4.** Using van't hoff equation, average desorption temperature of adsorbed hydrogen molecules are 305 K.

Strong binding of Sc is analyzed by computing the PDOS, Bader charge transfer, and visualizing the charge density difference plot. PDOS shows that the reason behind the strong bonds is the charge transfer of 3d orbital of Sc atom to C 2p orbitals of the BPh sheet. Bader charge analysis predicts that 1.6e charge gets transferred. The binding between $H_2$ and BPh+Sc are due to charge transfer from 3d orbital of Sc atom to $H_2$ molecules and back donation of charge from the $H_2$ to the transition metal 3d orbital, leading to a Kubas interaction. The presence of an energy barrier between possible diffusion paths prevents metal-metal clustering. Furthermore, structure stability is verified by AIMD at room and high temperatures by considering different configurations such as BPh+Sc, BPh+4Sc, and BPh+8Sc. Hence, we hope a computationally predicted stable Sc decorated BPh sheet with properties such as high gravimetric weight percentage and near ambient temperature desorption would be a potential candidate for solid-state hydrogen storage and draw the attention of experimentalists to further verification of its capacity in laboratory and industry.

**Acknowledgment**

MS sincerely wants to thank CSIR for supporting his Ph.D. research projects. MS also appreciate the Spacetime High Performance Computing team and its facility at IIT Bombay for their assistance. Drs. T. Shakuntala and Nandini Garg are acknowledged by BC for their encouragement and assistance. BC also thanks Drs. S.M. Yusuf and A. K. Mohanty for their assistance.

**References:**
[1] EIA projects 28% increase in world energy use by 2040 n.d.
[2] Renewable energy explained - U.S. Energy Information Administration (EIA) n.d.
[3] Shafiee S, Topal E. When will fossil fuel reserves be diminished? Energy Policy 2009;37:181–9.
[4] DOE Technical Targets for Hydrogen Production from Biomass-Derived Liquid Reforming. EnergyGov n.d.

[5]   DOE Technical Targets for Hydrogen Production from Electrolysis. EnergyGov n.d.
[6]   DOE Technical Targets for Hydrogen Production from Thermochemical Water Splitting. EnergyGov n.d.
[7]   DOE Technical Targets for Hydrogen Production from Photoelectrochemical Water Splitting. EnergyGov n.d.
[8]   DOE Technical Targets for Photobiological Hydrogen Production. EnergyGov n.d.
[9]   DOE Technical Targets for Hydrogen Production from Microbial Biomass Conversion. EnergyGov n.d.
[10]  DOE Technical Targets for Hydrogen Production from Biomass Gasification. EnergyGov n.d.
[11]  Hydrogen Storage. EnergyGov n.d.
[12]  DOE Technical Targets for Onboard Hydrogen Storage for Light-Duty Vehicles. EnergyGov n.d.
[13]  Usman MR. Hydrogen storage methods: Review and current status. Renewable and Sustainable Energy Reviews 2022:11.
[14]  Allendorf MD, Hulvey Z, Gennett T, Ahmed A, Autrey T, Camp J, et al. An assessment of strategies for the development of solid-state adsorbents for vehicular hydrogen storage. Energy Environ Sci 2018;11:2784–812.
[15]  Hassan IA, Ramadan HS, Saleh MA, Hissel D. Hydrogen storage technologies for stationary and mobile applications: Review, analysis and perspectives. Renewable and Sustainable Energy Reviews 2021;149:111311.
[16]  Andersson J, Grönkvist S. Large-scale storage of hydrogen. International Journal of Hydrogen Energy 2019;44:11901–19.
[17]  Dewangan SK, Mohan M, Kumar V, Sharma A, Ahn B. A comprehensive review of the prospects for future hydrogen storage in materials-application and outstanding issues. International Journal of Energy Research 2022;n/a.
[18]  Chen Z, Kirlikovali KO, Idrees KB, Wasson MC, Farha OK. Porous materials for hydrogen storage. Chem 2022;8:693–716.
[19]  Boateng E, Chen A. Recent advances in nanomaterial-based solid-state hydrogen storage. Materials Today Advances 2020;6:100022.
[20]  Chen Z, Ma Z, Zheng J, Li X, Akiba E, Li H-W. Perspectives and challenges of hydrogen storage in solid-state hydrides. Chinese Journal of Chemical Engineering 2021;29:1–12.
[21]  de Castro JFR, Santos SF, Costa ALM, Yavari AR, Botta F WJ, Ishikawa TT. Structural characterization and dehydrogenation behavior of Mg–5 at.%Nb nano-composite processed by reactive milling. Journal of Alloys and Compounds 2004;376:251–6.
[22]  Liang G. Synthesis and hydrogen storage properties of Mg-based alloys. Journal of Alloys and Compounds 2004;370:123–8.
[23]  Ponthieu M, Fernández JF, Cuevas F, Bodega J, Ares JR, Adeva P, et al. Thermodynamics and reaction pathways of hydrogen sorption in Mg6(Pd,TM) (TM = Ag, Cu and Ni) pseudo-binary compounds. International Journal of Hydrogen Energy 2014;39:18291–301.
[24]  Jain IP, Lal C, Jain A. Hydrogen storage in Mg: A most promising material. International Journal of Hydrogen Energy 2010;35:5133–44.
[25]  Wang Y, Adroher XC, Chen J, Yang XG, Miller T. Three-dimensional modeling of hydrogen sorption in metal hydride hydrogen storage beds. Journal of Power Sources 2009;194:997–1006.
[26]  CONDON JB. SURFACE AREA AND POROSITY DETERMINATIONS BY PHYSISORPTION: measurements and theory. Place of Publication Not Identified: ELSEVIER; 2019.
[27]  Sule R, Mishra AK, Nkambule TT. Recent advancement in consolidation of MOFs as absorbents for hydrogen storage. International Journal of Energy Research 2021;45:12481–99.
[28]  Rosi NL. Hydrogen Storage in Microporous Metal-Organic Frameworks. Science 2003;300:1127–9.

[29] Chen Z, Li P, Anderson R, Wang X, Zhang X, Robison L, et al. Balancing volumetric and gravimetric uptake in highly porous materials for clean energy. Science 2020;368:297–303.

[30] Langmi HW, Walton A, Al-Mamouri MM, Johnson SR, Book D, Speight JD, et al. Hydrogen adsorption in zeolites A, X, Y and RHO. Journal of Alloys and Compounds 2003;356–357:710–5.

[31] Deniz CU. Computational screening of zeolite templated carbons for hydrogen storage. Computational Materials Science 2022;202:110950.

[32] Kundu A, Trivedi R, Garg N, Chakraborty B. Novel permeable material "yttrium decorated zeolite templated carbon" for hydrogen storage: Perspectives from density functional theory. International Journal of Hydrogen Energy 2022.

[33] Li J, Furuta T, Goto H, Ohashi T, Fujiwara Y, Yip S. Theoretical evaluation of hydrogen storage capacity in pure carbon nanostructures. J Chem Phys 2003;119:2376–85.

[34] Ma L-P, Wu Z-S, Li J, Wu E-D, Ren W-C, Cheng H-M. Hydrogen adsorption behavior of graphene above critical temperature. International Journal of Hydrogen Energy 2009;34:2329–32.

[35] Srinivas G, Zhu Y, Piner R, Skipper N, Ellerby M, Ruoff R. Synthesis of graphene-like nanosheets and their hydrogen adsorption capacity. Carbon 2010;48:630–5.

[36] Aboutalebi SH, Aminorroaya-Yamini S, Nevirkovets I, Konstantinov K, Liu HK. Enhanced Hydrogen Storage in Graphene Oxide-MWCNTs Composite at Room Temperature. Advanced Energy Materials 2012;2:1439–46.

[37] Chandrakumar KRS, Ghosh SK. Alkali-Metal-Induced Enhancement of Hydrogen Adsorption in C60 Fullerene: An ab Initio Study. Nano Lett 2008;8:13–9.

[38] Sun Q, Jena P, Wang Q, Marquez M. First-Principles Study of Hydrogen Storage on Li12C60. J Am Chem Soc 2006;128:9741–5.

[39] Yoon M, Yang S, Hicke C, Wang E, Geohegan D, Zhang Z. Calcium as the Superior Coating Metal in Functionalization of Carbon Fullerenes for High-Capacity Hydrogen Storage. Phys Rev Lett 2008;100:206806.

[40] Yoon M, Yang S, Wang E, Zhang Z. Charged Fullerenes as High-Capacity Hydrogen Storage Media. Nano Lett 2007;7:2578–83.

[41] Wang Q, Sun Q, Jena P, Kawazoe Y. Theoretical Study of Hydrogen Storage in Ca-Coated Fullerenes. J Chem Theory Comput 2009;5:374–9.

[42] Zhao Y, Kim Y-H, Dillon AC, Heben MJ, Zhang SB. Hydrogen Storage in Novel Organometallic Buckyballs. Phys Rev Lett 2005;94:155504.

[43] Shin WH, Yang SH, Goddard WA, Kang JK. Ni-dispersed fullerenes: Hydrogen storage and desorption properties. Appl Phys Lett 2006;88:053111.

[44] Chakraborty B, Modak P, Banerjee S. Hydrogen Storage in Yttrium-Decorated Single Walled Carbon Nanotube. J Phys Chem C 2012;116:22502–8.

[45] Yildirim T, Ciraci S. Titanium-Decorated Carbon Nanotubes as a Potential High-Capacity Hydrogen Storage Medium. Phys Rev Lett 2005;94:175501.

[46] Shalabi AS, Abdel Aal S, Assem MM, Abdel Halim WS. Ab initio characterization of Ti decorated SWCNT for hydrogen storage. International Journal of Hydrogen Energy 2013;38:140–52.

[47] Tada K, Furuya S, Watanabe K. Ab initio study of hydrogen adsorption to single-walled carbon nanotubes. Phys Rev B 2001;63:155405.

[48] Ataca C, Aktürk E, Ciraci S, Ustunel H. High-capacity hydrogen storage by metallized graphene. Appl Phys Lett 2008;93:043123.

[49] Park N, Hong S, Kim G, Jhi S-H. Computational Study of Hydrogen Storage Characteristics of Covalent-Bonded Graphenes. J Am Chem Soc 2007;129:8999–9003.

[50] Ao ZM, Peeters FM. High-capacity hydrogen storage in Al-adsorbed graphene. Phys Rev B 2010;81:205406.

[51] Ao ZM, Jiang Q, Zhang RQ, Tan TT, Li S. Al doped graphene: A promising material for hydrogen storage at room temperature. Journal of Applied Physics 2009;105:074307.

[52] Yadav A, Chakraborty B, Gangan A, Patel N, Press MR, Ramaniah LM. Magnetic Moment Controlling Desorption Temperature in Hydrogen Storage: A Case of Zirconium-Doped Graphene as a High Capacity Hydrogen Storage Medium. J Phys Chem C 2017;121:16721–30.

[53] Zou J, Liu Z, Ge Y, Ding J, Nie M, Miao Z, et al. DFT study on hydrogen storage of Be or V modified boron-doped porous graphene. Materials Science in Semiconductor Processing 2022;150:106884.

[54] Cid BJ, Sosa AN, Miranda Á, Pérez LA, Salazar F, Mtz-Enriquez AI, et al. Enhanced reversible hydrogen storage performance of light metal-decorated boron-doped siligene: A DFT study. International Journal of Hydrogen Energy 2022.

[55] Chen I-N, Wu S-Y, Chen H-T. Hydrogen storage in N- and B-doped graphene decorated by small platinum clusters: A computational study. Applied Surface Science 2018;441:607–12.

[56] Faye O, Eduok U, Szpunar J, Szpunar B, Samoura A, Beye A. Hydrogen storage on bare Cu atom and Cu-functionalized boron-doped graphene: A first principles study. International Journal of Hydrogen Energy 2017;42:4233–43.

[57] Wang Y, Meng Z, Liu Y, You D, Wu K, Lv J, et al. Lithium decoration of three dimensional boron-doped graphene frameworks for high-capacity hydrogen storage. Appl Phys Lett 2015;106:063901.

[58] Gangan A, Chakraborty B, Ramaniah LM, Banerjee S. First principles study on hydrogen storage in yttrium doped graphyne: Role of acetylene linkage in enhancing hydrogen storage. International Journal of Hydrogen Energy 2019;44:16735–44.

[59] Gao Y, Zhang H, Pan H, Li Q, Zhao J. Ultrahigh hydrogen storage capacity of holey graphyne. Nanotechnology 2021;32:215402.

[60] Kubas GJ. Hydrogen activation on organometallic complexes and H2 production, utilization, and storage for future energy. Journal of Organometallic Chemistry 2009;694:2648–53.

[61] Wang L, Yang RT. New sorbents for hydrogen storage by hydrogen spillover – a review. Energy Environ Sci 2008;1:268.

[62] Modak P, Chakraborty B, Banerjee S. Study on the electronic structure and hydrogen adsorption by transition metal decorated single wall carbon nanotubes. J Phys: Condens Matter 2012;24:185505.

[63] Cho ES, Ruminski AM, Aloni S, Liu Y-S, Guo J, Urban JJ. Graphene oxide/metal nanocrystal multilaminates as the atomic limit for safe and selective hydrogen storage. Nat Commun 2016;7:10804.

[64] Gu J, Zhang X, Fu L, Pang A. Study on the hydrogen storage properties of the dual active metals Ni and Al doped graphene composites. International Journal of Hydrogen Energy 2019;44:6036–44.

[65] Tarasov BP, Arbuzov AA, Mozhzhuhin SA, Volodin AA, Fursikov PV, Lototskyy MV, et al. Hydrogen storage behavior of magnesium catalyzed by nickel-graphene nanocomposites. International Journal of Hydrogen Energy 2019;44:29212–23.

[66] Samantaray SS, Sangeetha V, Abinaya S, Ramaprabhu S. Enhanced hydrogen storage performance in Pd3Co decorated nitrogen/boron doped graphene composites. International Journal of Hydrogen Energy 2018;43:8018–25.

[67] Luo Y, Ren C, Xu Y, Yu J, Wang S, Sun M. A first principles investigation on the structural, mechanical, electronic, and catalytic properties of biphenylene. Sci Rep 2021;11:19008.

[68] Denis PA, Iribarne F. Hydrogen storage in doped biphenylene based sheets. Computational and Theoretical Chemistry 2015;1062:30–5.

[69] Mahamiya V, Shukla A, Chakraborty B. Ultrahigh reversible hydrogen storage in K and Ca decorated 4-6-8 biphenylene sheet. International Journal of Hydrogen Energy 2022:S0360319922004578.

[70] Mane P, Kaur SP, Chakraborty B. Enhanced reversible hydrogen storage efficiency of zirconium-decorated biphenylene monolayer: A computational study. Energy Storage n.d.;n/a:e377.

[71] Malyi OI, Sopiha KV, Persson C. Energy, Phonon, and Dynamic Stability Criteria of Two-Dimensional Materials. ACS Appl Mater Interfaces 2019;11:24876–84.

[72] Kresse G, Furthmüller J. Efficiency of ab-initio total energy calculations for metals and semiconductors using a plane-wave basis set. Computational Materials Science 1996;6:15–50.

[73] Kresse G, Furthmüller J. Efficient iterative schemes for *ab initio* total-energy calculations using a plane-wave basis set. Phys Rev B 1996;54:11169–86.

[74] Perdew JP, Burke K, Ernzerhof M. Generalized Gradient Approximation Made Simple. Phys Rev Lett 1996;77:3865–8.

[75] Perdew JP, Chevary JA, Vosko SH, Jackson KA, Pederson MR, Singh DJ, et al. Atoms, molecules, solids, and surfaces: Applications of the generalized gradient approximation for exchange and correlation. Phys Rev B 1992;46:6671–87.

[76] Fuchs M, Bockstedte M, Pehlke E, Scheffler M. Pseudopotential study of binding properties of solids within generalized gradient approximations: The role of core-valence exchange correlation. Phys Rev B 1998;57:2134–45.

[77] Chakraborty B, Ray P, Garg N, Banerjee S. High capacity reversible hydrogen storage in titanium doped 2D carbon allotrope Ψ-graphene: Density Functional Theory investigations. International Journal of Hydrogen Energy 2021;46:4154–67.

[78] Grimme S. Semiempirical GGA-type density functional constructed with a long-range dispersion correction. Journal of Computational Chemistry 2006;27:1787–99.

[79] Monkhorst HJ, Pack JD. Special points for Brillouin-zone integrations. Phys Rev B 1976;13:5188–92.

[80] Wang V, Xu N, Liu J-C, Tang G, Geng W-T. VASPKIT: A user-friendly interface facilitating high-throughput computing and analysis using VASP code. Computer Physics Communications 2021;267:108033.

[81] Togo A, Tanaka I. First principles phonon calculations in materials science. Scripta Materialia 2015;108:1–5.

[82] Zhang Y, Cheng X. Hydrogen storage property of alkali and alkaline-earth metal atoms decorated C24 fullerene: A DFT study. Chemical Physics 2018;505:26–33.

[83] Kresse G, Hafner J. Ab initio molecular-dynamics simulation of the liquid-metal--amorphous-semiconductor transition in germanium. Phys Rev B 1994;49:14251–69.

[84] Kresse G, Hafner J. Ab initio molecular dynamics for liquid metals. Phys Rev B 1993;47:558–61.

[85] Nosé S. A unified formulation of the constant temperature molecular dynamics methods. J Chem Phys 1984;81:511–9.

[86] Fan Q, Yan L, Tripp MW, Kachel SR, Chen M, Foster AS, et al. Biphenylene network: A nonbenzenoid carbon allotrope 2021:6.

[87] Bafekry A, Faraji M, Fadlallah MM, Jappor HR, Karbasizadeh S, Ghergherehchi M, et al. Biphenylene monolayer as a two-dimensional nonbenzenoid carbon allotrope: a first-principles study. J Phys: Condens Matter 2022;34:015001.

[88] Wang J, Ma F, Sun M. Graphene, hexagonal boron nitride, and their heterostructures: properties and applications. RSC Adv 2017;7:16801–22.

[89] Yee Chung SY, Tomita M, Yokogawa R, Ogura A, Watanabe T. Atomic mass dependency of a localized phonon mode in SiGe alloys. AIP Advances 2021;11:115225.

[90] Vankayala RK, Lan T-W, Parajuli P, Liu F, Rao R, Yu SH, et al. High zT and Its Origin in Sb-doped GeTe Single Crystals. Advanced Science 2020;7:2002494.

[91] Fung V, Ganesh P, Sumpter BG. Physically Informed Machine Learning Prediction of Electronic Density of States. Chem Mater 2022;34:4848–55.

[92] Arnaldsson A, Tang W, Chill S, Chai W, Henkelman G. bader charge analysis n.d.

[93] Kubas GJ. Metal–dihydrogen and σ-bond coordination: the consummate extension of the Dewar–Chatt–Duncanson model for metal–olefin π bonding. Journal of Organometallic Chemistry 2001;635:37–68.

[94] Kubas GJ. Activation of dihydrogen and coordination of molecular H2 on transition metals. Journal of Organometallic Chemistry 2014;751:33–49.

[95] Anikina EV, Banerjee A, Beskachko VP, Ahuja R. Influence of Kubas-type interaction of B–Ni codoped graphdiyne with hydrogen molecules on desorption temperature and storage efficiency. Materials Today Energy 2020;16:100421.

[96] Varunaa R, Ravindran P. Potential hydrogen storage materials from metal decorated 2D-$C_2N$: an *ab initio* study. Phys Chem Chem Phys 2019;21:25311–22.

[97] Jain V, Kandasubramanian B. Functionalized graphene materials for hydrogen storage. J Mater Sci 2020;55:1865–903.

[98] Anikina E, Banerjee A, Beskachko V, Ahuja R. Li-decorated carbyne for hydrogen storage: charge induced polarization and van't Hoff hydrogen desorption temperature. Sustainable Energy Fuels 2020;4:691–9.

[99] Hu S, Yong Y, Li C, Zhao Z, Jia H, Kuang Y. $Si_2BN$ monolayers as promising candidates for hydrogen storage. Phys Chem Chem Phys 2020;22:13563–8.

[100] Durgun E, Ciraci S, Yildirim T. Functionalization of carbon-based nanostructures with light transition-metal atoms for hydrogen storage. Phys Rev B 2008;77:085405.

[101] Haynes WM, editor. CRC Handbook of Chemistry and Physics. 97th ed. Boca Raton: CRC Press; 2016.

[102] Fair KM, Cui XY, Li L, Shieh CC, Zheng RK, Liu ZW, et al. Hydrogen adsorption capacity of adatoms on double carbon vacancies of graphene: A trend study from first principles. Phys Rev B 2013;87:014102.

[103] Kittel C. Introduction to solid state physics. 8. ed., [repr.]. Hoboken, NJ: Wiley; 20.

**Figures and tables:**

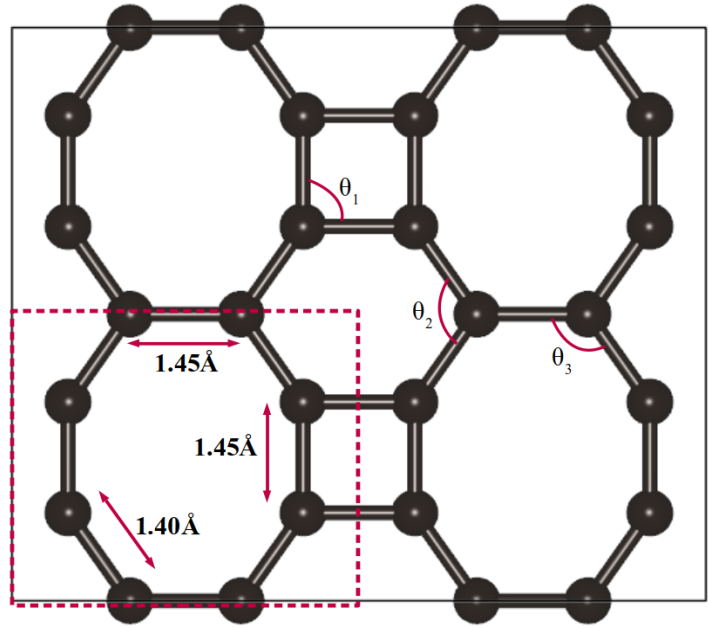

**Fig. 1:** Geometrically optimized 2x2x1 supercell of BPh sheet. The magenta color inset represents the unit cell of the BPh sheet. Double headed arrows show bond-lengths, and the bond angles are $\theta_1=90°$, $\theta_2=109.89°$, $\theta_3=125.05°$.

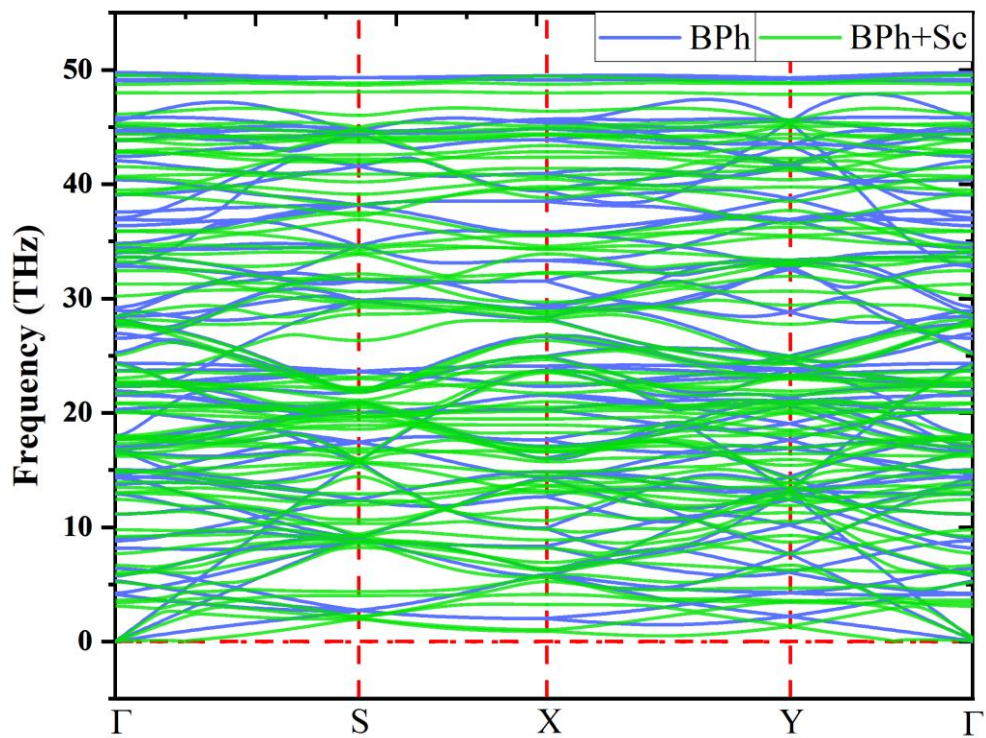

**Fig. 2:** Phonon spectrum of BPh sheet (blue color) and Sc decorated BPh sheet (green color). Heavy atom Sc (compared with C-atom) doping shifted the phonon-spectrum lower near fermi-level and highest optical modes (50 THz).

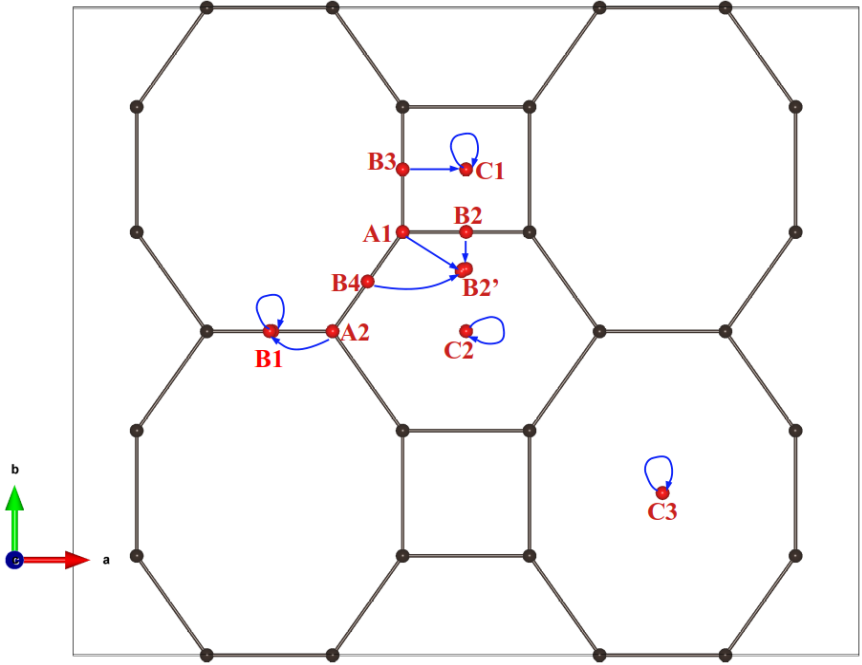

**Fig. 3:** A schematic diagram showing all possible chosen positions for doping and optimized Sc atoms on BPh sheet. The tail and head of blue arrows show the position of decorated Sc atoms before and after optimization. The Black and red spheres represent carbon and scandium atoms, respectively.

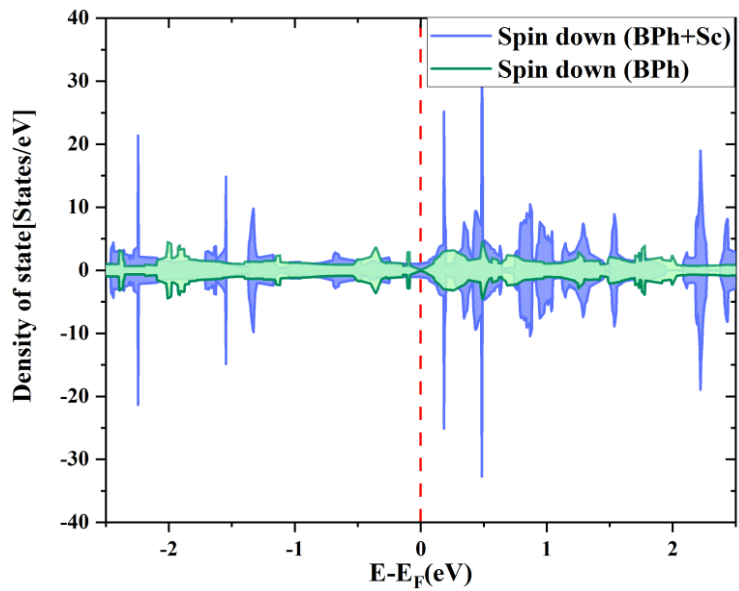

**Fig. 4:** Total density of states of BPh sheet before (green) and after (blue) decorating Sc atom on it.

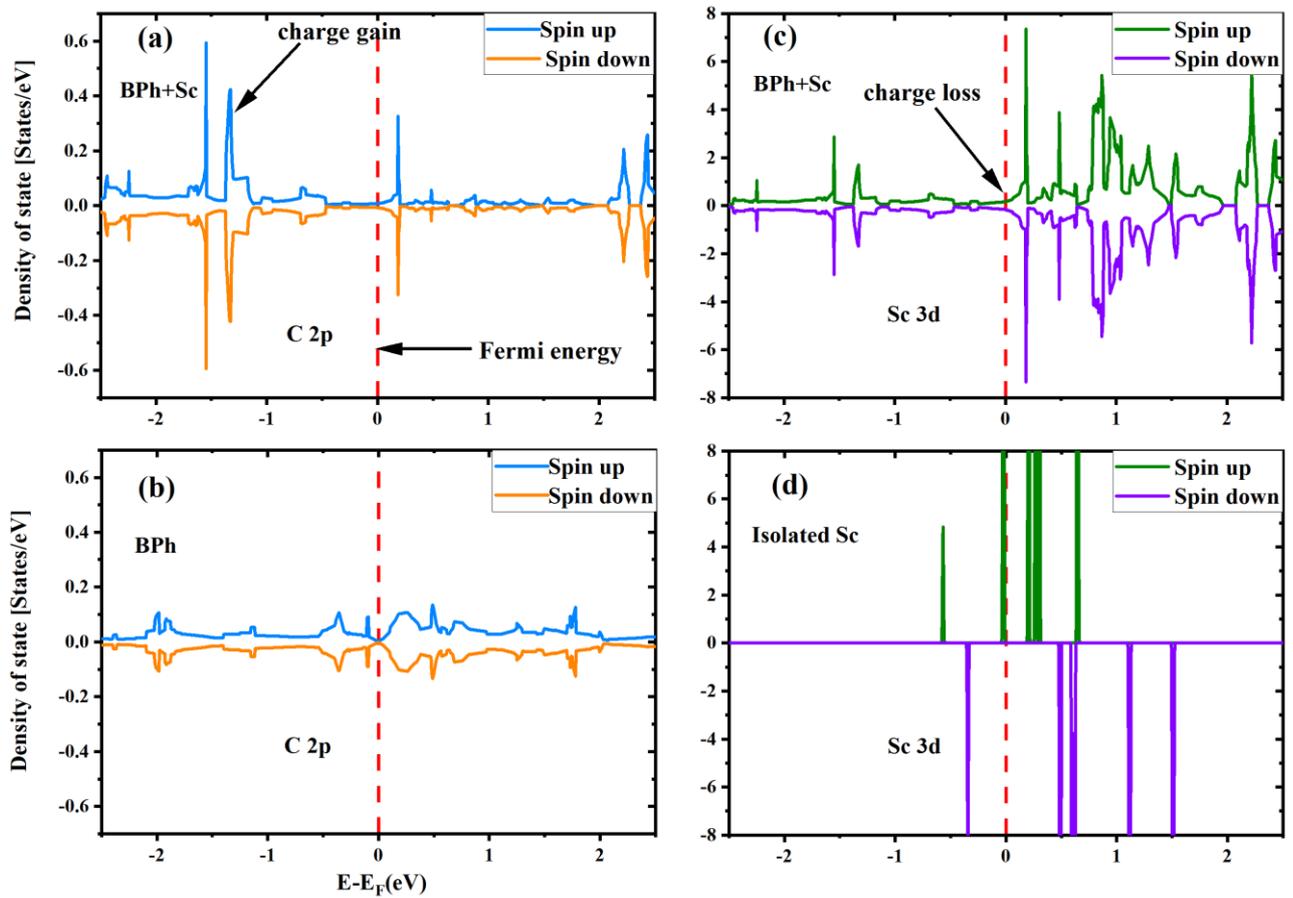

**Fig. 5:** Partial density of states (PDOS) for interaction of Sc with BPh. In the left panel, PDOS of the 2p orbital of carbon atom of (a) BPh+Sc and (b) BPh are given. In the right panel, PDOS of 3d orbital of Sc atom for (c) BPh+Sc (d) isolated Sc atom are given.

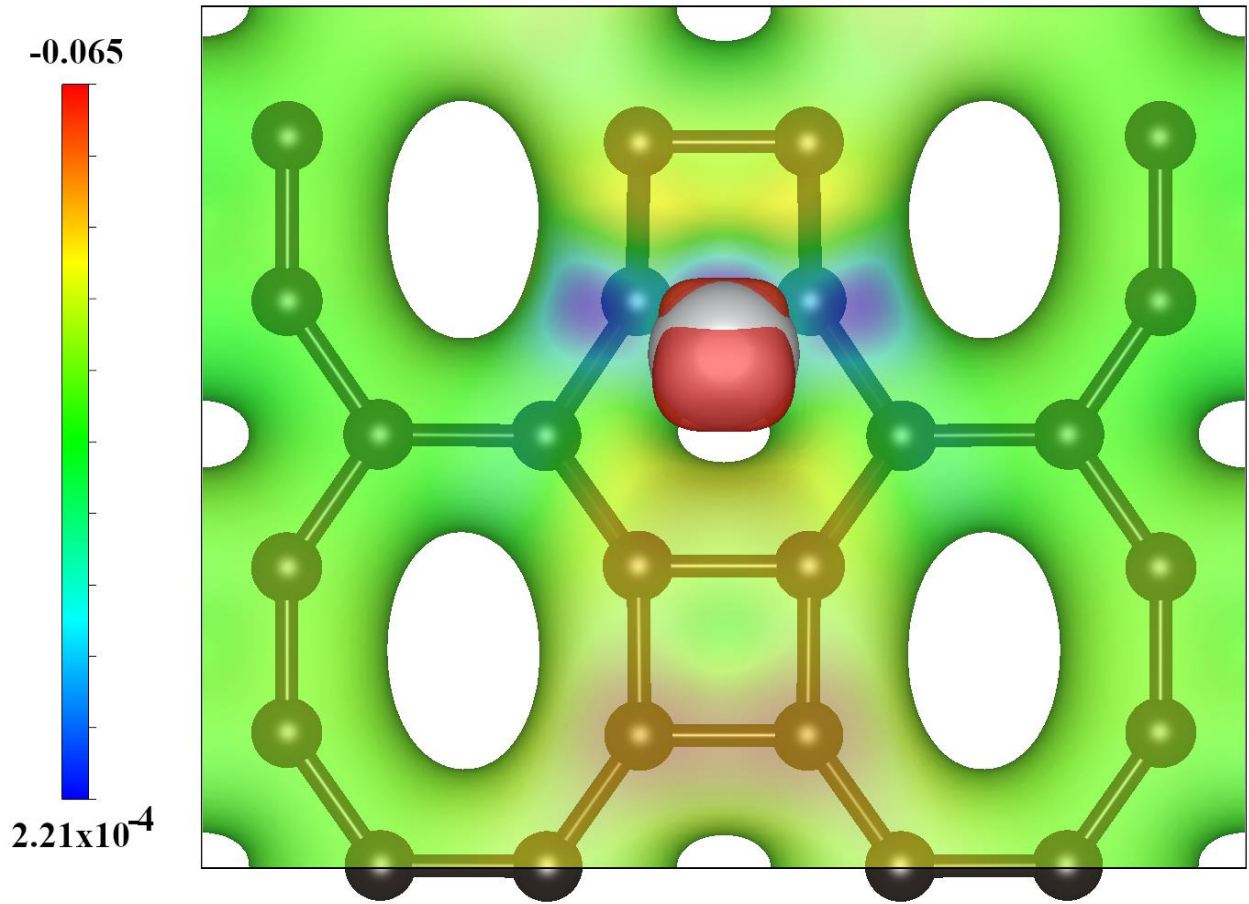

**Fig. 6:** Difference of charge density of BPh+Sc and BPh sheet, i.e., ρ(BPh+Sc@B2')-ρ(BPh) with isovalue $4.23 \times 10^{-2}$, to visualize spatial charge transfer between Sc and BPh sheet. The red and Blue regions show charge depletion and gain regions, respectively.

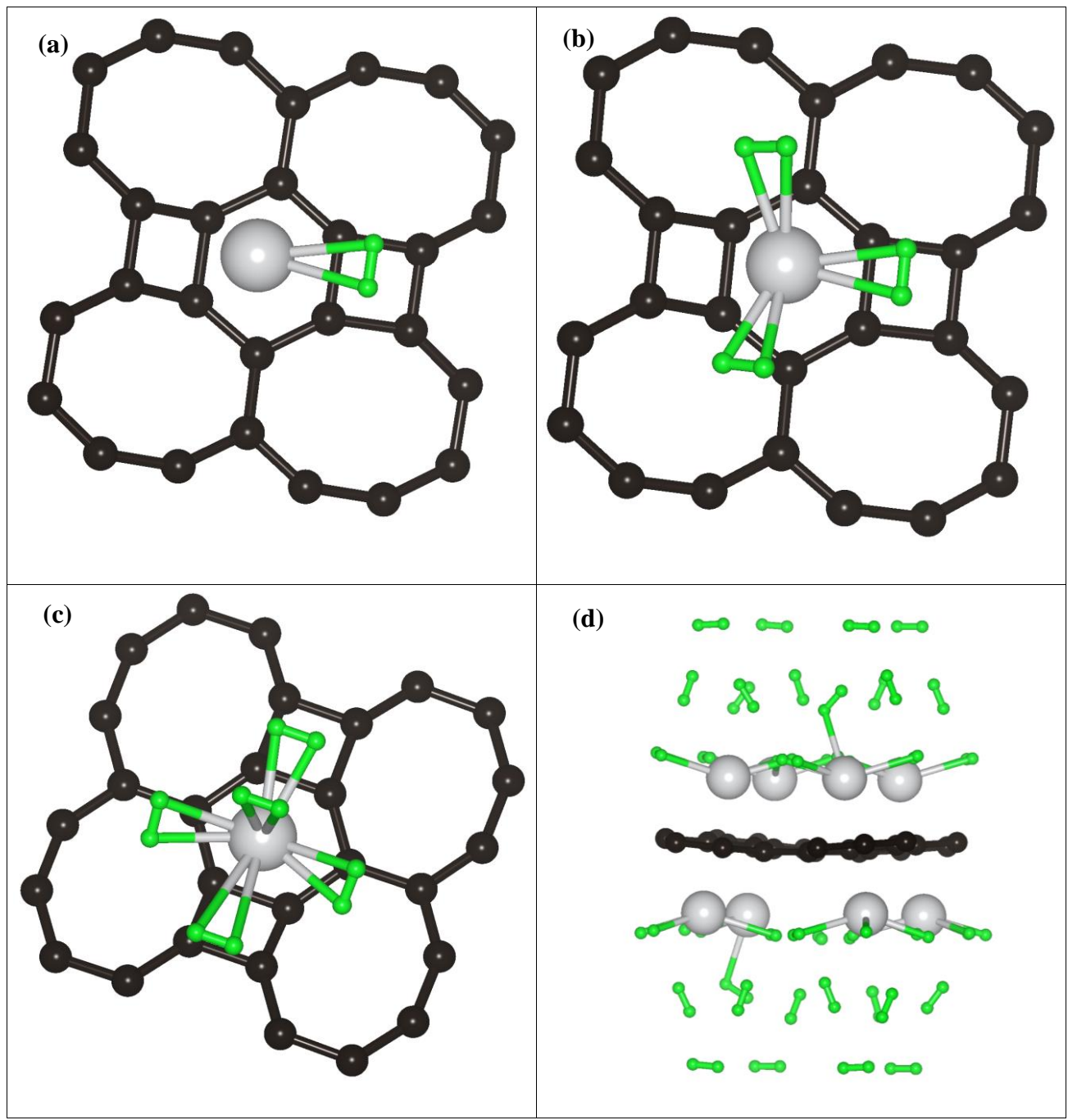

**Fig. 7:** DFT+Grimme-D2 optimized structure of n-$H_2$ adsorbed Sc decorated BPh. (a) BPh+Sc+1$H_2$ (b) BPh+Sc+3$H_2$ (c) BPh+Sc+5$H_2$ (d) BPh+8Sc+40$H_2$.

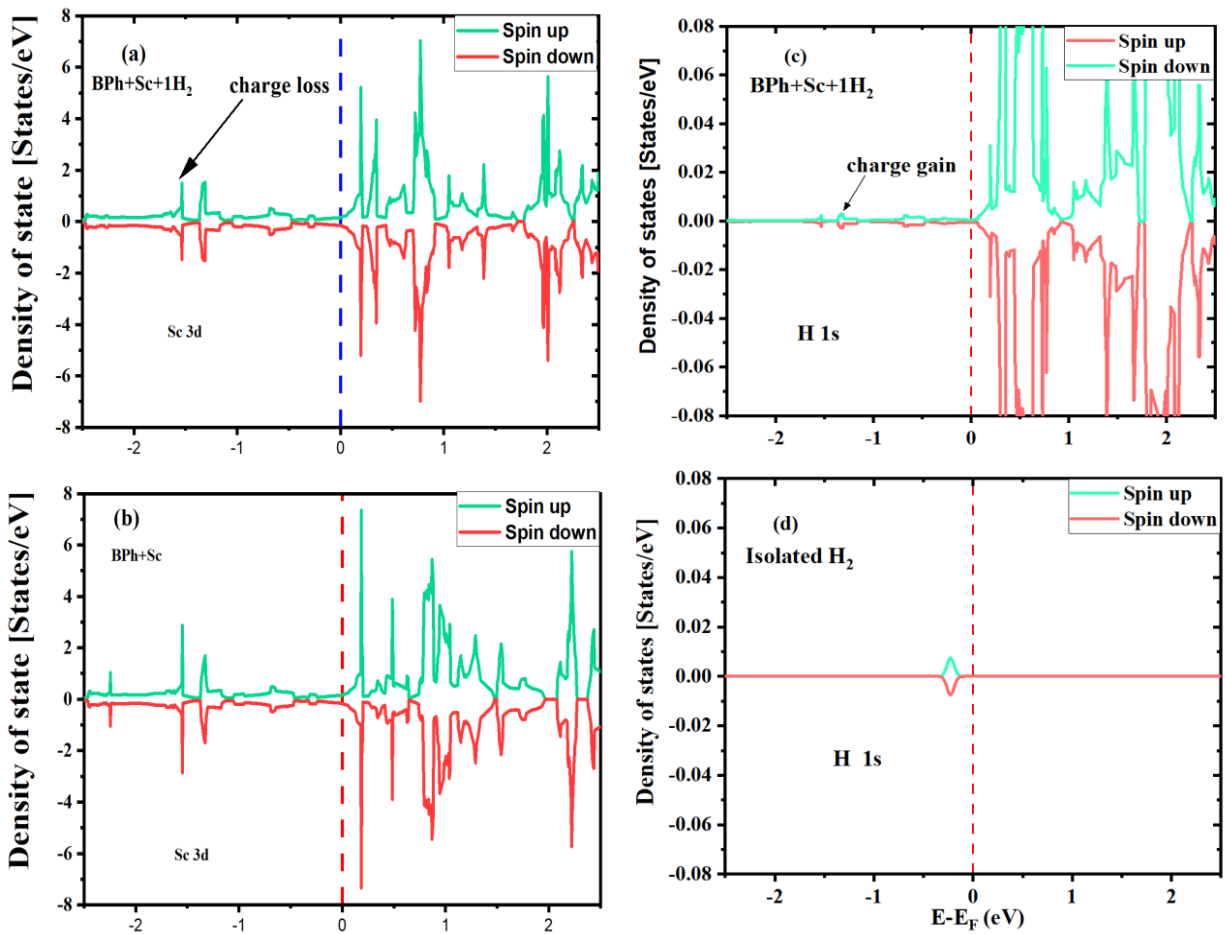

**Fig. 8:** PDOS for interaction of $H_2$ molecules with Sc+BPh system. PDOS of 3d orbital of Sc atom of **(a)** BPh+Sc+1$H_2$ **(b)** BPh+Sc; PDOS of 1s orbital of **(c)** BPh+Sc+1$H_2$ and **(d)** isolated $H_2$ molecule.

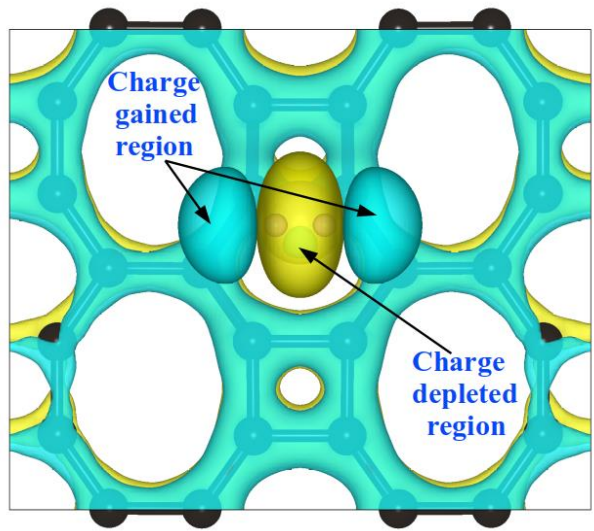

**Fig. 9:** Charge density difference of BPh+Sc+1H$_2$ and BPh+Sc and isolated H$_2$ molecule i.e., ρ(BPh+Sc+1H$_2$)-ρ(BPh+Sc)-ρ(H$_2$) with isovalue 3.5x10$^{-3}$e.

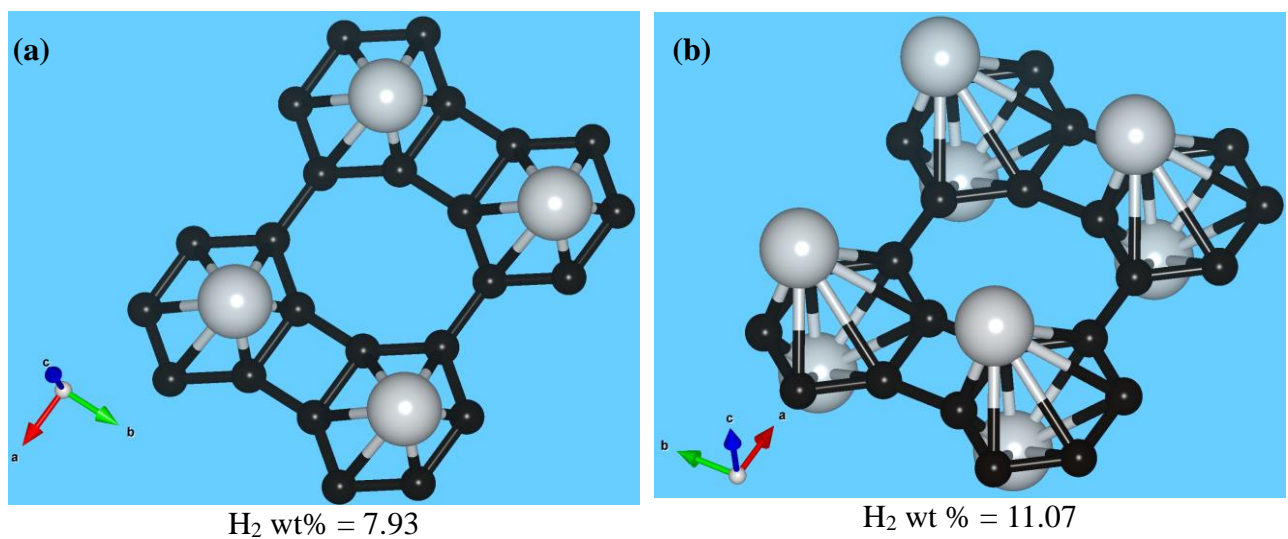

H$_2$ wt% = 7.93    H$_2$ wt % = 11.07

**Fig. 10:** Metal loading configurations and their corresponding gravimetric hydrogen percentage of : (a) Sc doped on one side of BPh sheet (b) Sc doped on both sides of BPh sheet.

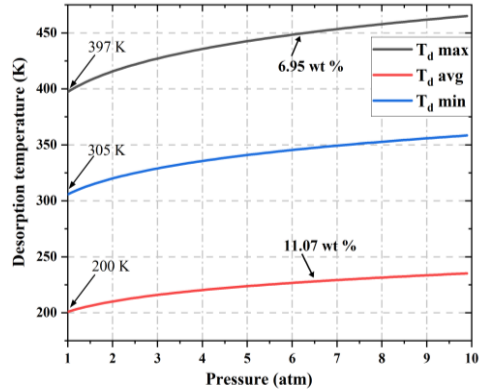

**Fig. 11:** Dependence of desorption temperature of BPh+Sc system with binding energy and equilibrium pressure. The highest, lowest and average adsorption energy of $H_2$ corresponds to highest, lowest and average gravimetric wt% represented with blue, red and magenta color respectively.

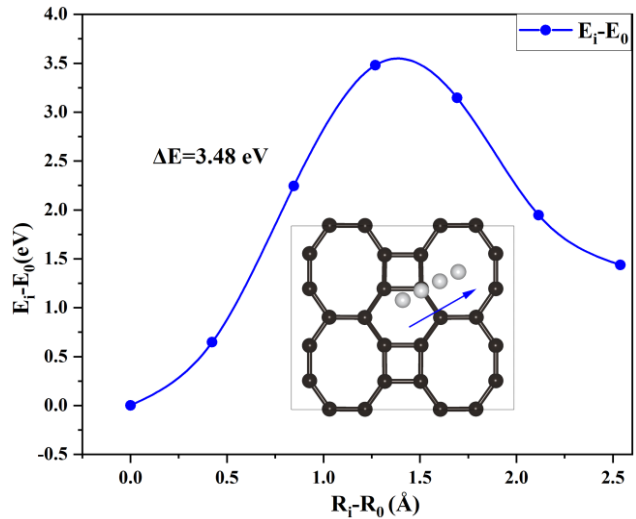

**Fig. 12:** Diffusion energy barrier between B2' and C3 positions. The energy barrier are found to be 3.5 eV, enough to prevent diffusion.

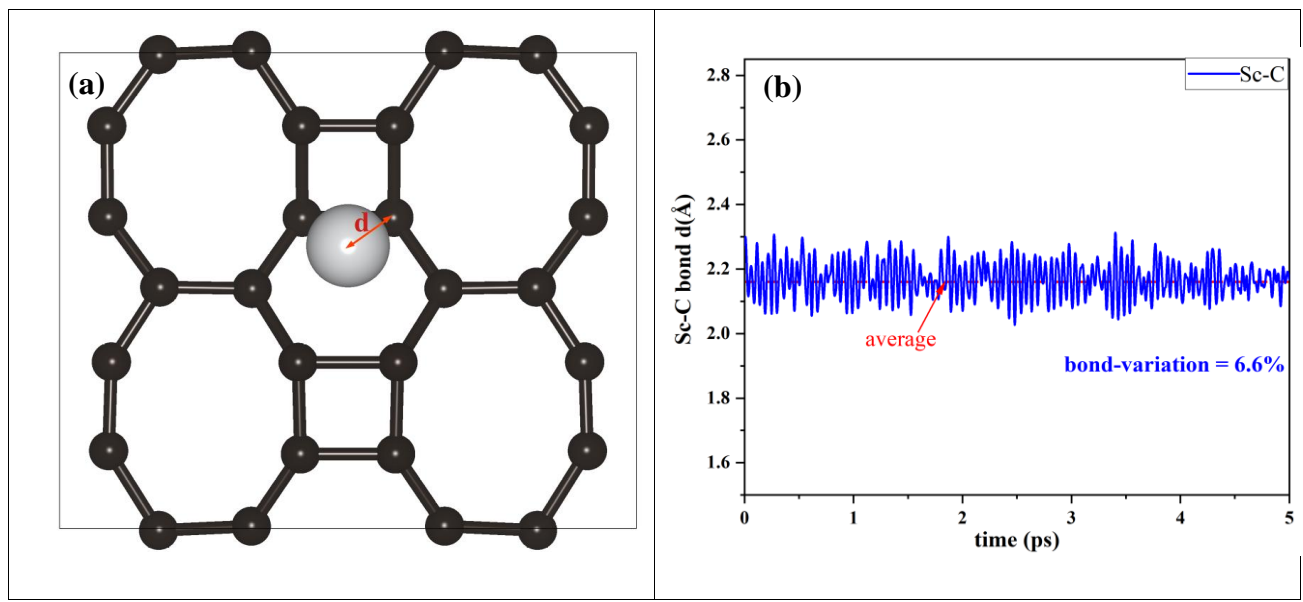

**Fig. 13:** Ab-initio molecular dynamics of (a) snapshot of BPh+Sc at 300K (b) the variation of bond distance between Sc and its nearest C atom.

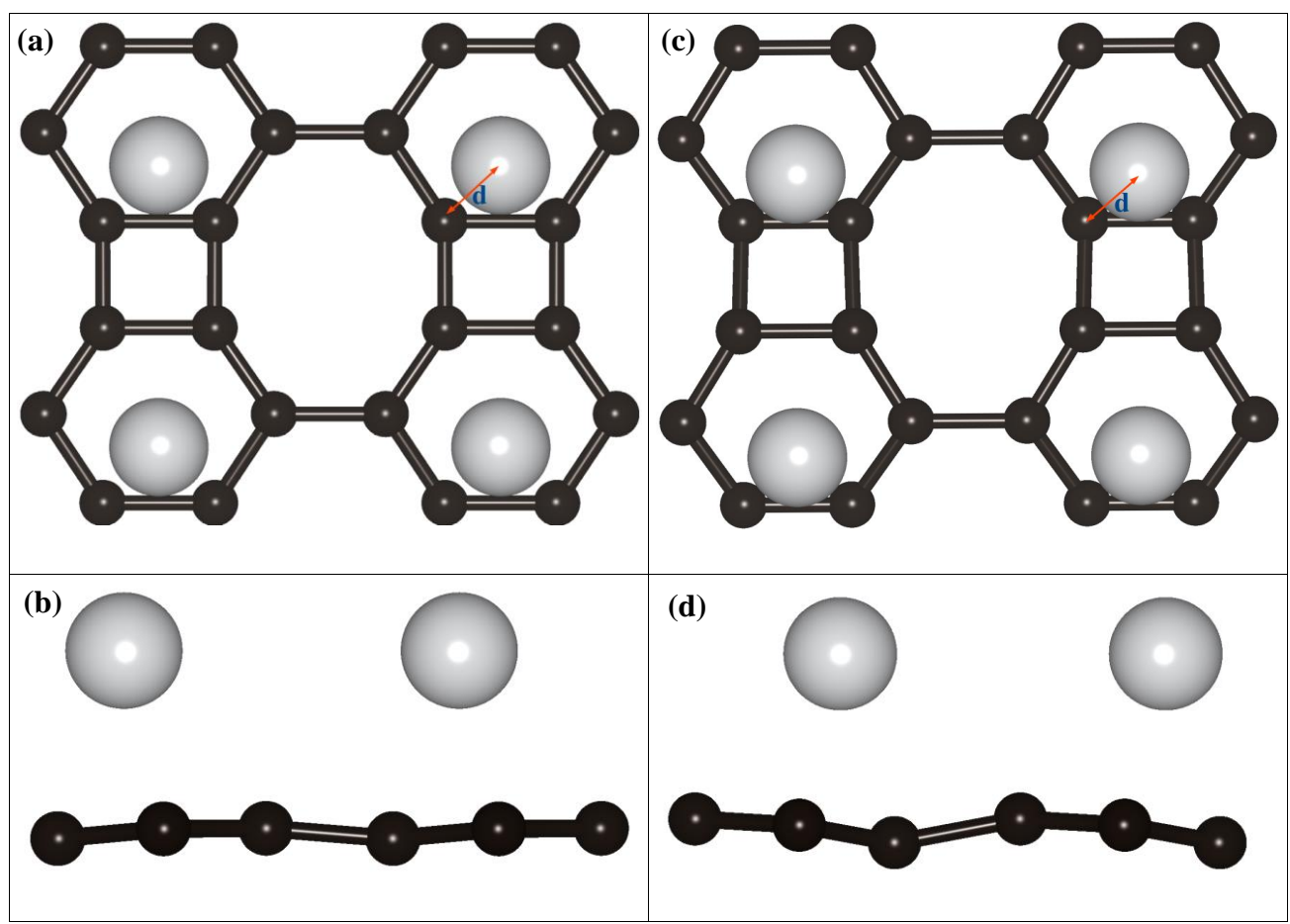

**Fig. 14:** Top and side view of AIMD simulation for Sc atom decorated on each hexagonal site at (a) 300K top view (b) 300K side view (c) the higher temperature (~700 K) top view (d) ~700 K side view, which shows the system are stable at room and higher temperature without clustering.

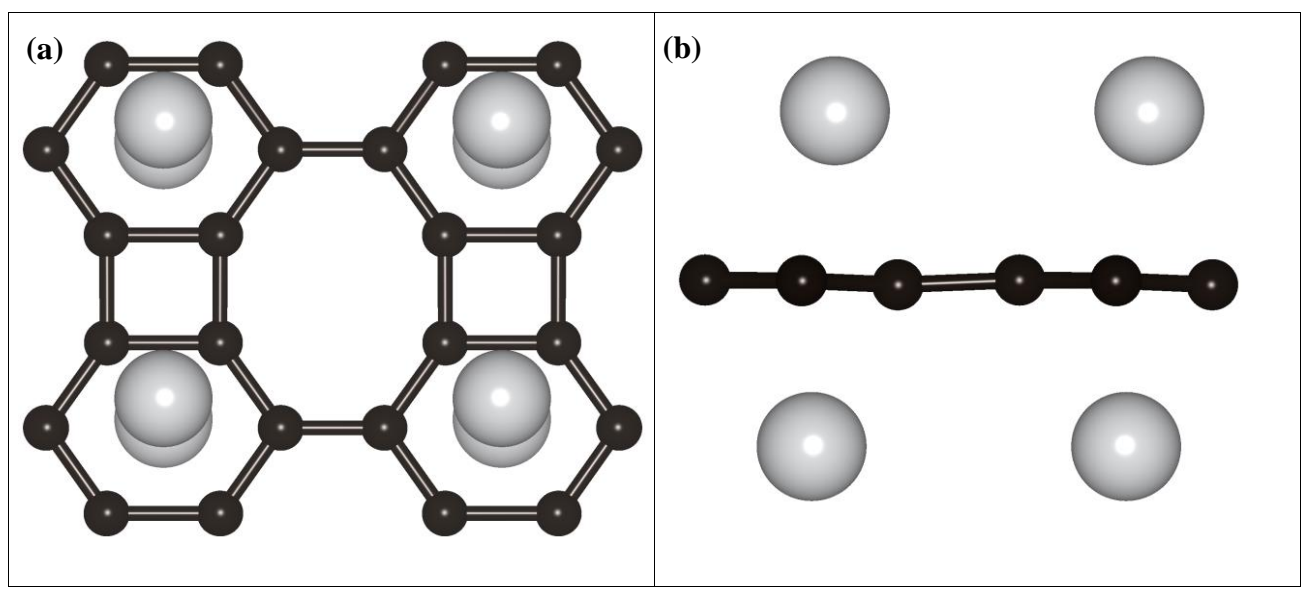

**Fig. 15:** Top (a) and side (b) view of AIMD of eight Sc atoms decorated at every hexagonal site on both side, which finally confirm the stability of BPh+8Sc system with no clustering.

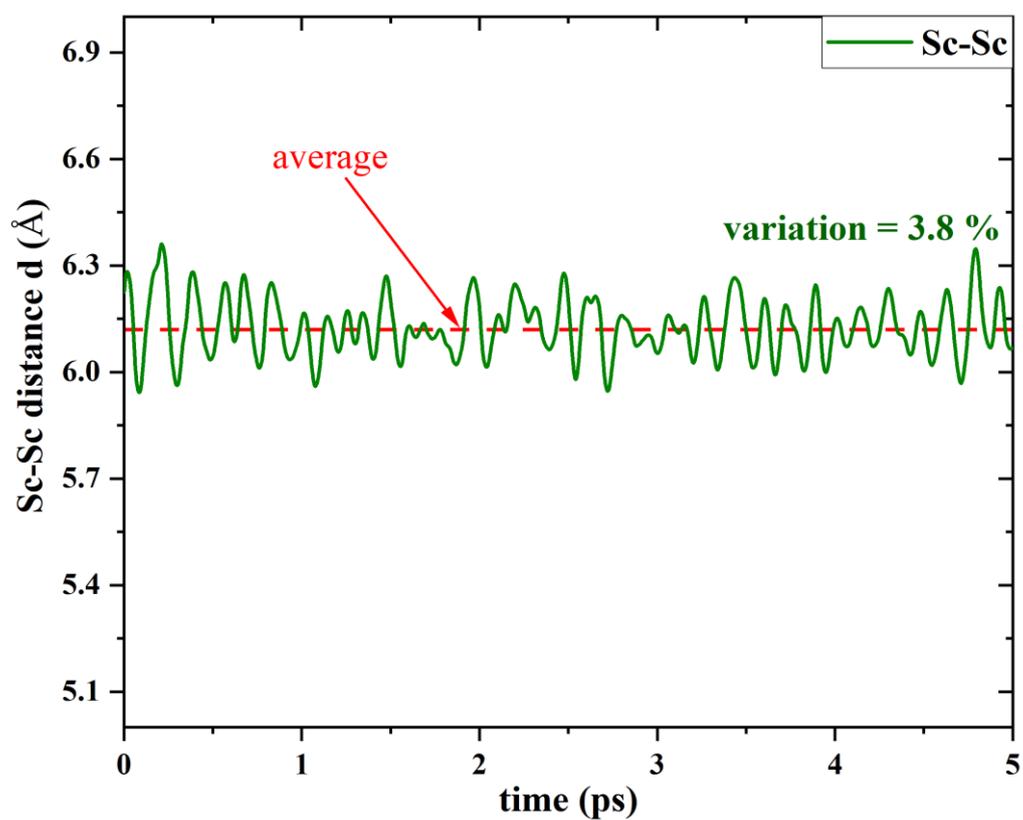

**Fig. 16:** The variation of distance between two Sc of BPh+4Sc(@B2') at 300K.

**Table 1:** The binding energy of decorated Sc atom at various positions on BPH along with their distances between Sc to its nearest carbon of BPH and Bader charge transfer from Sc to BPH sheet.

| S.N | System | Distance (Sc-C) Å | Binding energy(eV) | Bader charge donated from Sc |
|---|---|---|---|---|
| 1. | BPh+Sc@B1 | 2.162 | -2.90 | 1.26e |
| 2. | BPh+Sc@B2' | 2.154 | -3.83 | 1.34e |
| 3. | BPh+Sc@C1 | 2.188 | -3.22 | 1.29e |
| 4. | BPh+Sc@C2 | 2.223 | -3.56 | 1.32e |
| 5. | BPh+Sc@C3 | 2.302 | -2.90 | 1.36e |

**Table 2:** The adsorption energy of n-$H_2$ molecules on Sc decorated BPH using GGA (PBE), and GGA+ Grimme-D2 corrected functional:

| S.N. | System | GGA+DFT-D2(eV) | H-H(Å) | Sc-H(Å) |
|---|---|---|---|---|
| 1. | BPh+Sc | -4.046 | | |
| 2. | BPh+Sc+1$H_2$ | -0.428 | 0.760 | 2.423 |
| 3. | BPh+Sc+3$H_2$ | -0.538 | 0.779 | 2.220 |
| 4. | BPh+Sc+5$H_2$ | -0.272 | 0.752 | 3.682 |
| Average binding energy per $H_2$ | | -0.412 eV | | |

**Table 3:** Comparison the hydrogen storage efficiency of our system with theoretical and experimental studies of other 2D systems:

| S.N. | Systems | DFT Codes-functionals | $H_2$ gravimetric wt% |
|---|---|---|---|
| 1. | Li+graphene [48] | VASP-LDA | 12.8 |
| 2. | Al-adsorbed graphene [50] | DMOL3-LDA | 13.79 |
| 3. | Zr+graphene [52] | VASP-GGA+DFT-D2 | 11.0 |
| 4. | Be modified B-doped porous | VASP-GGA(PBE)+DFT-D3 | 7.03 |
| | graphene [53] | | |
| 5. | Ca decorated B-doped siligene[54] | SIESTA-GGA(PBE)+DFT-D2 | 13.79 |
| 6. | Y+graphyne [58] | VASP-GGA+DFT-D2 | 10.0 |
| 7. | Li+Holey graphyene [59] | VASP-GGA(PBE)+DFT-D2 | 12.8 |
| 8. | Ti+Ψ-graphene [77] | VASP-GGA+DFT-D3 | 13.14 |
| **9.** | **BPh+Sc [current work]** | **VASP-GGA(PBE)+DFT-D2** | **11.07** |
| | Experimental works | | |
| 9. | Mg nanocrystal encaptulated in graphene [63] | | 6.5 |
| 10. | Ni and Al codoped graphene [64] | | 5.7 |
| 11. | Ni, Mg composite doped on graphene [65] | | > 6.5 |
| 12. | $Pd_3Co$ decorated boron doped graphene [66] | | ~4.5 |

# Highly Efficient Hydrogen Storage of Sc Decorated Biphenylene Monolayer near Ambient-temperature: An Ab-initio Simulations

*Mukesh Singh[a], Alok Shukla[a], Brahmananda Chakraborty[b,c,*]*

[a]Department of Physics, Indian Institute of Technology Bombay, Powai, Mumbai 400076, India

[b]High Pressure and Synchrotron Radiation Physics Division, Bhabha Atomic Research Centre, Trombay, Mumbai, India

[c]Homi Bhabha National Institute, Mumbai, India

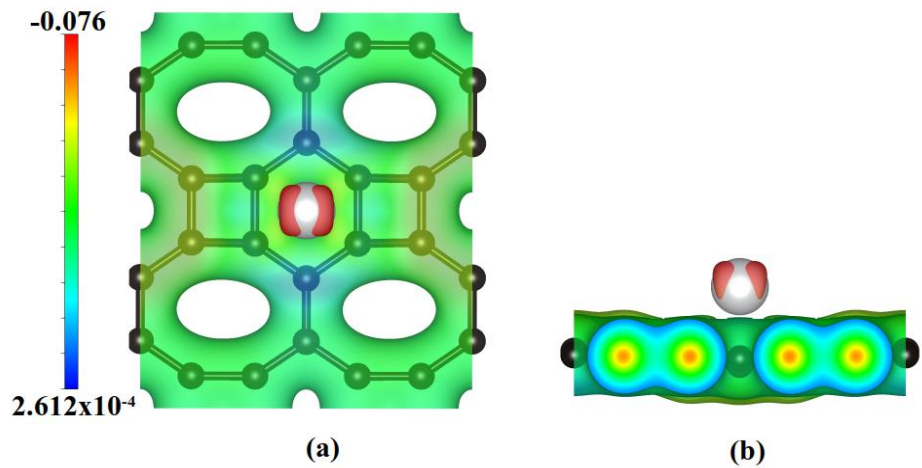

**Fig. S 1:** Difference of charge density of BPh+Sc and BPh sheet, ρ(BPh+Sc@C2)-ρ(BPh) with isovalue 4.54x10$^{-2}$ (a) top view (b) side view.

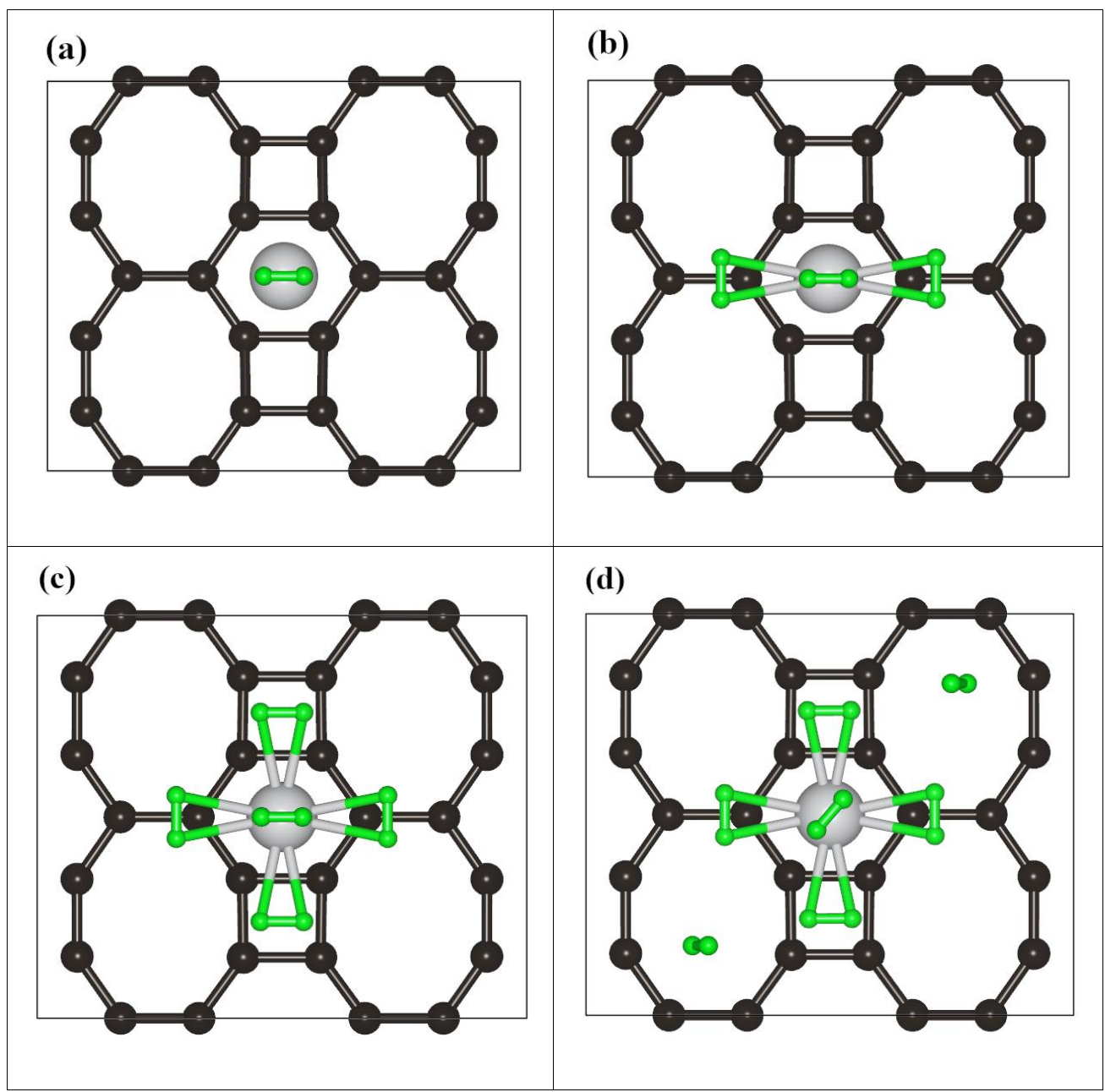

**Fig. S2:** DFT+Grimme-D2 optimized structure of nH$_2$ adsorbed on BPh+Sc(@C2).
(a) BPh+Sc+1H$_2$ (b) BPh+Sc+3H$_2$ (c) Bph+Sc+5H$_2$ (d) BPh+Sc+7H$_2$.

**Table S1:** The adsorption energy of nH$_2$ molecules on BPh+Sc(@C2) using GGA (PBE) and GGA+ Grimme-D2 corrected functional:

| S.N. | System | GGA(eV) | GGA+DFT-D2(eV) |
|---|---|---|---|
| 1. | BPh+Sc | -3.563 | -3.811 |
| 2. | BPh+Sc+1H$_2$ | -0.353 | -0.399 |
| 3. | BPh+Sc+3H$_2$ | -0.379 | -0.492 |
| 4. | BPh+Sc+5H$_2$ | -0.314 | -0.453 |
| 5. | BPh+Sc+7H$_2$ | -0.125 | -0.143 |
| Average binding energy per H$_2$ (up to 5H$_2$) | | -0.293 eV | -0.448 eV |

Since 7$^{th}$ H$_2$ molecule binding energy is not in the range of DoE range (0.2-0.7), we consider only five doped hydrogen molecules adsorbed on Sc+BPh(@C2), and their corresponding gravimetric weight percentage turns out to be:

     **one side**  (24C+4Sc+20H$_2$)  :    **7.93 wt%**
     **both side** (24C+8Sc+40H$_2$)  :    **11.07 wt%**

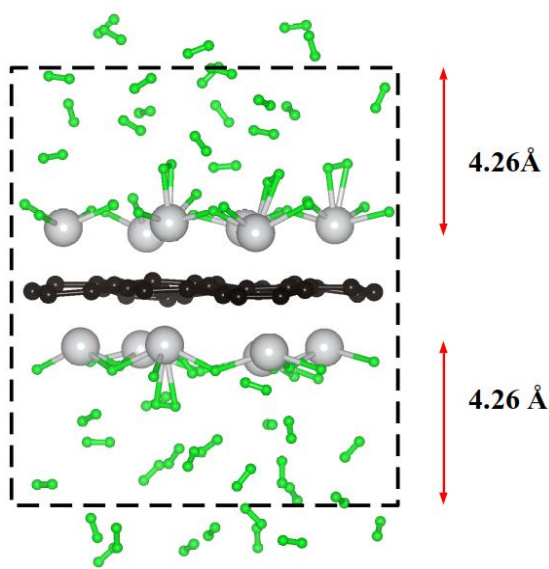

**Fig. S3:** Side view of BPh+8Sc+40H$_2$ after 5ps AIMD simulations performed at room temperature.

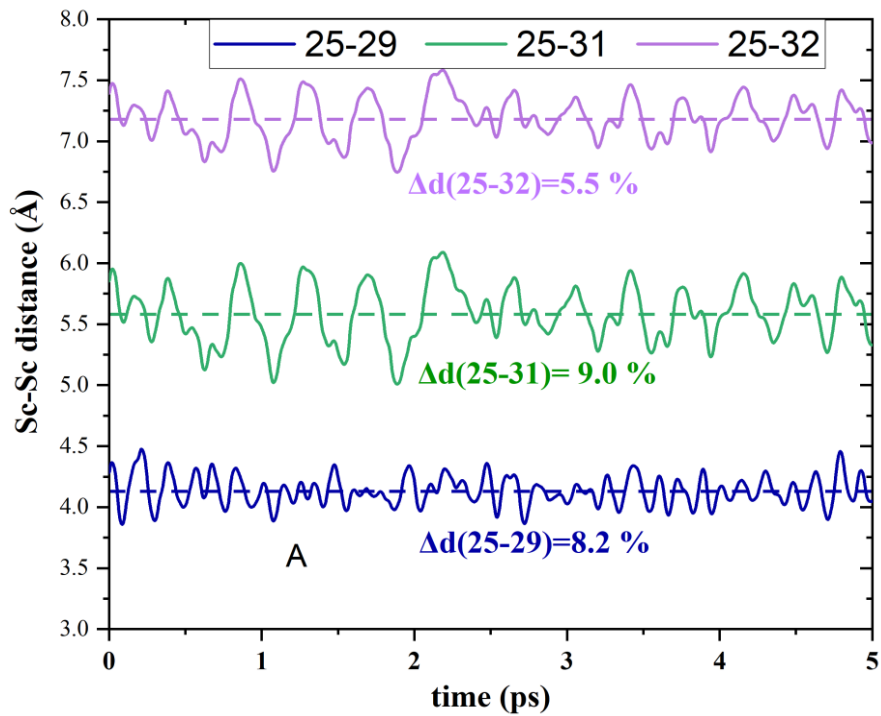

**Fig. S4:** The variation of distance between two Sc of BPh+8Sc(@B2') at 300K. Δd(N1, N2) denotes the maximum percentage deviation w.r.t. mean values between two $N_1$, $N_2^{th}$ Sc atoms. Dotted lines show the average of distance.

## Doping of Sc atoms at different sites simultaneously on BPh sheet

**1-** After decorating Sc atoms at all porous positions i.e. at the center of square, hexagon and octagon, we performed geometry relaxation, MD-simulations and plotted the final structures in Figs. S4(b) and S4(c) respectively.

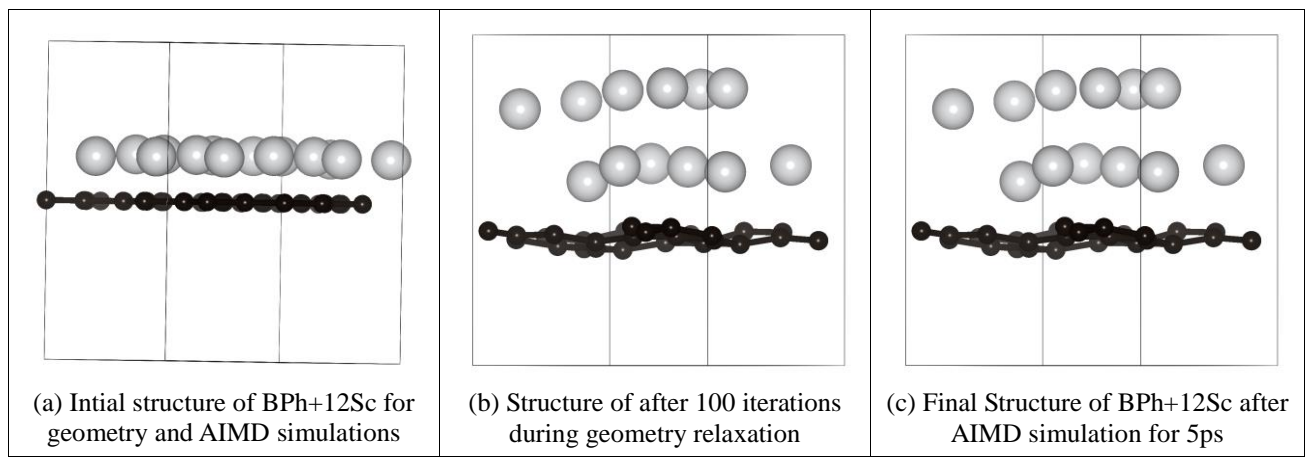

(a) Intial structure of BPh+12Sc for geometry and AIMD simulations

(b) Structure of after 100 iterations during geometry relaxation

(c) Final Structure of BPh+12Sc after AIMD simulation for 5ps

**Fig S5:** (a) Plotting of structure when Sc atoms are decorated at all porous sites i.e., square, hexagon, and octagon. (b) Geometrically relaxed structure up to ~100 iterations. Since the structure seemed disturbed, relaxation is terminated. (c) AIMD simulation of structure for 5ps with steps of 1fs at 300K.

From both the geometrical relaxation and MD-simulation, it is clear that doping of Sc atoms on all porous sites are not possible.

**2-** After the finding unstability of Sc atoms doping on at all porous positions (hexagon, octagon and square), we explored the possibility of stability on decoration of Sc at hexagon and octagon positions. We plotted the geometry relaxation S5(b) and MD-simulations results in Figs. S6(b) and (c), respectively.

**Geometrical relaxation and MD-simulations:**

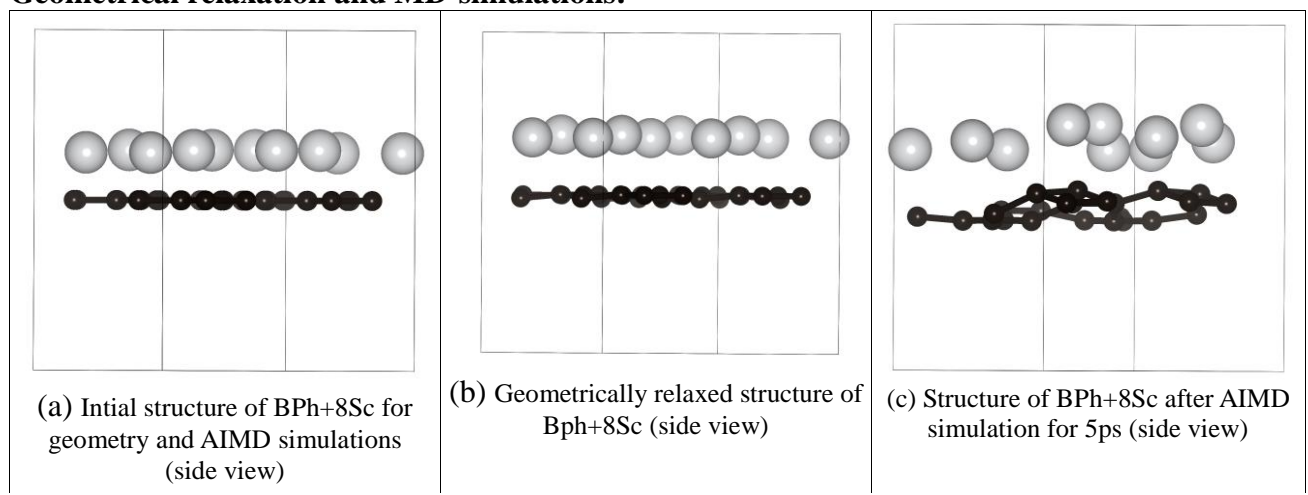

(a) Intial structure of BPh+8Sc for geometry and AIMD simulations (side view)

(b) Geometrically relaxed structure of Bph+8Sc (side view)

(c) Structure of BPh+8Sc after AIMD simulation for 5ps (side view)

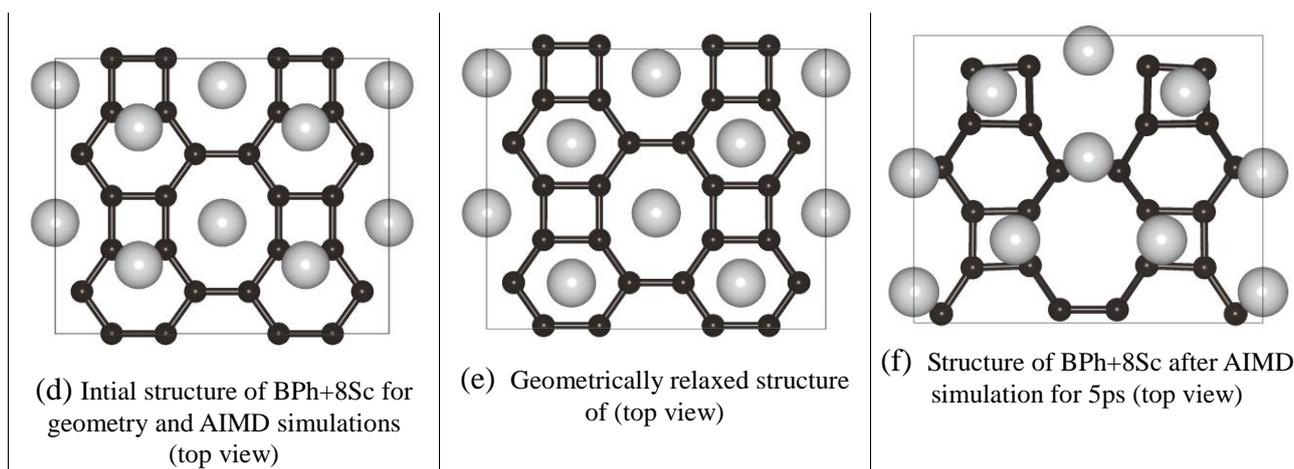

(d) Intial structure of BPh+8Sc for geometry and AIMD simulations (top view)

(e) Geometrically relaxed structure of (top view)

(f) Structure of BPh+8Sc after AIMD simulation for 5ps (top view)

**Fig S6:** (a,d) side and top view of BPh+8Sc when Sc atoms are decorated at hexagon and octagon sites (b,e) side and top view of the geometrically relaxed structure of BPh+8Sc. (c) AIMD simulation of structure for 5ps with steps of 1fs at 300K.

In geometrical relaxation of the system, all the Sc atoms are remain at the center of the hexagon and octagon. However, in AIMD-simulation, Sc atoms are displaced from the center to bond sites, i.e., AIMD-simulations exclude the possibility of simultaneous doping of Sc atoms at hexagon and octagon.

**3-** Being the octagon as the largest pore BPh sheet, one can think of doping Sc atoms at the center of octagon of BPh sheet. We have checked the binding energy of Sc atom by putting it at the center of octagons Fig. S6 and found the binding energy is positive (+3.89 eV), hence the Sc atom at the center of octagon is weakly bounded. To verify the weak strength of Sc at the center of octagon, we displaced Sc atom by 0.5 Å Fig. S7 (a) and after optimizaton, it moved over the center of octagon with negative binding energy (-2.9 eV) Fig. S7 (b) as in our previous manuscript for over center of octagon. We have repeated calculations with doping an Sc on each octagon and found similar results. Hence we conclude that the Sc doping at center of octagon is the local minima for BPh+Sc while over the center of octagon is stable.

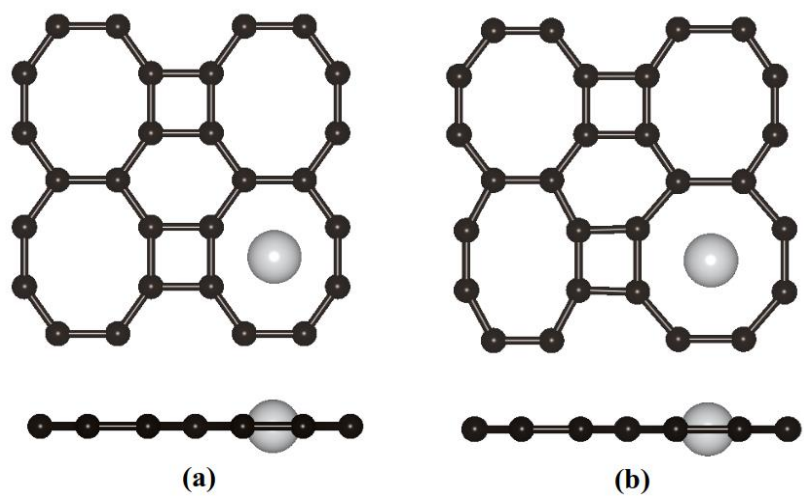

(a)         (b)

**Fig. S7:** Top and side view of initial (a) and final (b) optimized structure of Sc decorated BPh sheet.

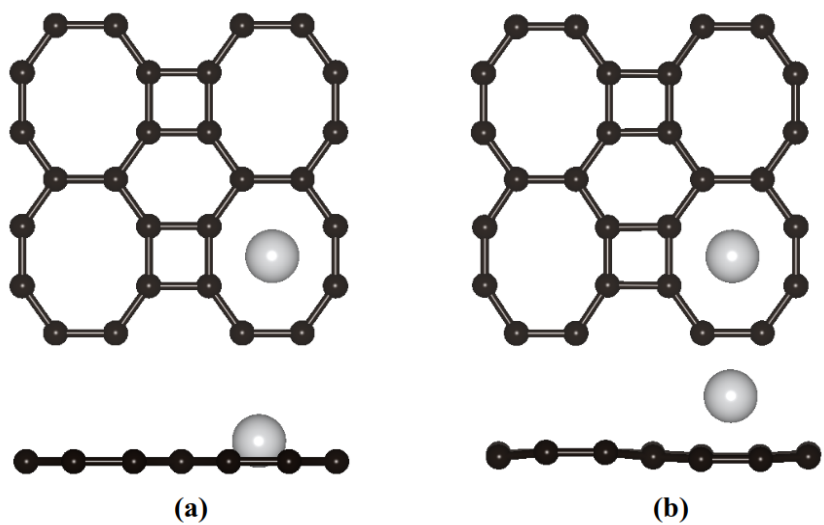

**Fig. S8:** Top and side view of initial strcuture (a) displaced by 0.5 Å from center of octagon and final (b) optimized structure of Sc decorated BPh sheet.

**H-H bond length of H$_2$ molecule:**

We have used the experimental bond length (0.74 Å) of the hydrogen molecule in our simulations for hydrogen storage. We have performed ground state energy calculations for different H-H separations starting from 0.70 Å to 0.80 Å as plotted below, which indicates that simulated H-H separation distance with PBE functional is 0.75Å.

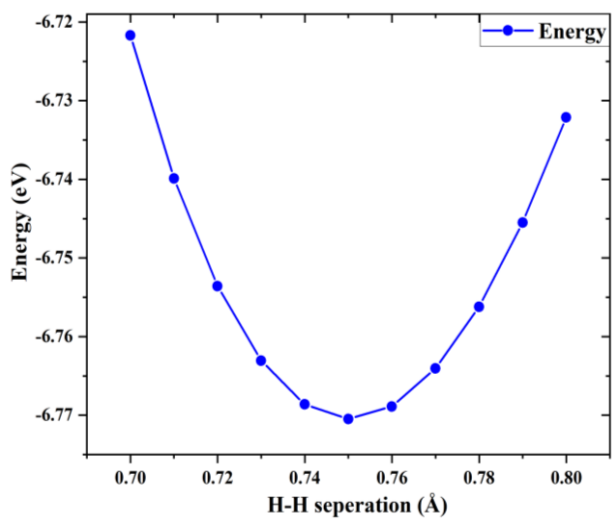

**Fig. S9:** DFT ground state energy with respect to H-H separation for H$_2$ molecules.